\documentclass{aa}  

\usepackage{txfonts}
\usepackage{amsmath}
\usepackage{amssymb}
\usepackage{mathrsfs}
\usepackage{courier}
\usepackage{titlesec}
\usepackage{graphicx}
\usepackage{float}
\usepackage{fancyvrb}
\usepackage{caption}
\usepackage{xcolor}
\usepackage{caption,subcaption}
\usepackage{geometry}
\usepackage{siunitx}
\usepackage{upgreek}
\begin{document} 

   \title{Spatial disconnection between stellar and dust emissions: the test of the Antennae Galaxies (Arp 244)}
   
   
   \author{L.-M. Seillé
          \inst{1}
          \and
          V. Buat\inst{1,2}
          \and
          W. Haddad\inst{1}
          \and
          A.Boselli\inst{1}
          \and
          M.Boquien\inst{3}
          \and
          L.Ciesla\inst{1}
          \and
          Y. Roehlly\inst{1}
          \and
          D.Burgarella\inst{1}
          }

   \institute{Aix Marseille Univ, CNRS, CNES, LAM, Marseille, France\\
              \email{lise-marie.seille@lam.fr}
         \and
             Institut Universitaire de France (IUF)\\
             \email{veronique.buat@lam.fr}
        \and
             Centro de Astronomía (CITEVA), Universidad de Antofagasta, Avenida Angamos 601, Antofagasta, Chile
             }

   \date{}

 \abstract
   {The detection with of the Atacama Large Millimeter Array (ALMA) of dust-rich high redshift galaxies whose cold dust emission is spatially disconnected from the ultraviolet emission bears a challenge for modelling their spectral energy distributions (SED) with codes based on an energy budget between the stellar and dust components.}
   {We want to test the validity of energy balance modelling on a nearby resolved galaxy with vastly different ultraviolet and infrared spatial distributions and infer what information can be reliably retrieved from the analysis of the full spectral energy distribution. }
   {We use 15 broadband images of the Antennae Galaxies ranging from far-ultraviolet to far-infrared and divide Arp 244 into 58 square \textasciitilde 1 kpc$^2$ regions. We fit the data with CIGALE to determine the star formation rate, stellar mass and dust attenuation of each region. We  compare these quantities to the ones obtained for Arp 244 as a whole.}
   {The  spectral energy distributions of the 58 regions and Arp 244 are well fitted. The estimates for the star formation rate and stellar mass for the whole galaxy and the addition of the 58 regions are found consistent within one sigma. We present the spatial distribution of these physical parameters as well as the shape of the attenuation curve across the Antennae Galaxies . We find that the Overlap Region exhibits a high star formation rate, attenuation and a shallow attenuation curve. We observe a flattening of the attenuation curves with increasing attenuation and dust surface densityin agreement with the predictions of hydrodynamical simulations coupled with radiative transfer modelling. }
   {}


   \maketitle
%

\section{Introduction}

Modelling and fitting the spectral energy distributions (SED) of galaxies is one of the most popular methods to derive crucial physical  parameters to study galaxy evolution such as the star formation rate (SFR) and the stellar mass. However, dust can drastically change the shape of the ultraviolet (UV) to near infrared (IR) SED as it absorbs and scatters photons, mostly at wavelengths shorter than  1 $\mu$m, and thermally emits the absorbed energy in the infrared (from $\sim$ 3  to $\sim$ 1000 $\mu$m). Physically-based SED modelling codes  take this effect into account by applying the energy balance principle: the energy radiated by dust corresponds to the energy of the absorbed stellar light e.g. \citep{dacunha,noll,carnall,boquien19}. 
 SED fitting codes  introduce an attenuation recipe  to redden the stellar continuum. It  mostly consists of  the attenuation curve measured  for starburst galaxies \citep{calzetti} or of the recipe of \cite{charlot}  with a power-law dependence of the effective attenuation with wavelength and a differential amount of attenuation for young and older stars.

Variations in attenuation laws in local and distant galaxies have  been observed e.g. \citep[LF17 hereafter]{kriek,buat12,battisti17,salim18,boquien22,lofaro}. The attenuation law is found to flatten out when the attenuation increases e.g. \citep{salmon,buat18,boquien22}. Radiation transfer models applied to simple geometries \citep{chevallard} or to hydrodynamical simulations \citep{roebuck,trayford} also predict similar trends. Both observational and theoretical studies lead to the conclusion that a single attenuation curve does not account for all star-forming galaxies and that more flexible curves are required \citep{salim20}.

Recent observations of high-redshift massive dusty galaxies with the Atacama Large Millimeter Array (ALMA) found  a very compact dust emission, a clumpy rest-frame UV emission distributed in clumps located far away from the compact dust emission e.g. \citep{elbaz,rujopakarn19}. These very different dust and stellar distributions raise the question of the validity of a local energy balance which is nevertheless still expected to be valid at a global scale. The way stellar and dust emissions are linked is expected to have  an impact on the physical parameters derived from the modelling of the global emission. It may be at least partially reflected by a variable effective attenuation curve as shown in \cite{buat19}. 

In this paper we use NGC 4038/ NGC 4039 (also known as Arp244 or the Antennae Galaxies) as a proxy to test the accuracy of SED modelling techniques based on energy conservation  in high redshift dusty galaxies with very complex geometry e.g. \citep{dunlop,elbaz}. 
The Antennae Galaxies are a pair of interacting galaxies in their second encounter \citep{zhang} although \cite{privon} point towards Arp 244 being at a slightly earlier stage in its merging process. They are currently going through a starburst phase in which the collision of clouds of gas and dust causes rapid star formation. Their proximity of 21 Mpc \citep{riess}, allows for a multi-wavelength UV to far-IR  study at a kpc scale. Arp 244 exhibits very different UV and far-IR spatial distribution like the dusty high-redshift galaxies observed with ALMA. Our estimation for the distance between the regions of intense UV and far-IR emissions is $\simeq 7$  kpc in Arp 244 which can be considered as typical in dusty high redshift star forming galaxies e.g. \citep{rujopakarn16,hodge,gomez,rujopakarn19}. This makes the Antennae Galaxies a very good candidate  to investigate the reliability of  both global and local (kpc scale)  measurements of physical quantities.

The paper is structured as follows: in Section 2, we introduce the multi-band data that we use in this study, and describe the multi-wavelength photometry of individual regions. 
Section 3 is dedicated to the SEDs modules. 
We present our results in Section 4 and discuss them in Section 5. Conclusions are given in Section 6.

\section{Photometry}

\subsection{Multi-wavelengths images}
We retrieved images of the Antennae Galaxies from the UV to the far-IR. Both FUV (1516 Å) and NUV (2267 Å) images come  from the GALEX Ultraviolet Atlas of Nearby Galaxies distributed by \cite{gil}. The images in the \emph{g},\emph{r}, \emph{i}, \emph{z}, and \emph{y} bands are obtained from the Pan-STARRS DR2 survey catalogue \citep{chambers}. The VISTA \emph{Ks} image obtained during the VHS survey by \cite{mcmahon} was retrieved from the ESO Science Archive Facility. The \emph{Spitzer} data consists of the four Infrared Array Camera \citep[IRAC,][]{fazio} bands and the 24 $\mu$m band from the Multiband Imaging Photometer for \emph{Spitzer} \citep[MIPS,][]{rieke}. These images, processed by \cite{bendo}, were retrieved from the \emph{Spitzer} Heritage Archive. We got user-provided data products of the Photodetector Array Camera and Spectrometer \citep[PACS,][]{klaas} 70 (PACS-blue) and 160 (PACS-red) $\mu$m maps from the \emph{Herschel} \citep{pilbratt} Science Archive. We retrieved the PACS images processed with the program package Scanamorphos \citep{roussel}. The main characteristics of the filters used as well as the spatial resolution of the images and their minimum signal to noise ratios are given in Table 1.

\begin{table*}
\centering
\begin{tabular}{|l|r|r|r|r|}
\hline
  \multicolumn{1}{|c|}{Filters} &
  \multicolumn{1}{c|}{$\lambda_{eff}$ ($\mu$m)} &
  \multicolumn{1}{c|}{Bandwidth ($\mu$m)} &
  \multicolumn{1}{c|}{Processed Pixel Size} &
  \multicolumn{1}{c|}{Minimum S/N}\\
\hline
  FUV & $0.155$ & $0.134-0.181$ & $1.5$ \arcsec & 7.3\\
  NUV & $0.230$ & $0.169-0.301$ & $1.5$ \arcsec & 6.4\\
  PS g & $0.481$ & $0.394-0.559$ & $0.25$ \arcsec & 4.0\\
  PS r & $0.616$ & $0.539-0.704$ & $0.25$ \arcsec & 5.6\\
  PS i & $0.750$ & $0.678-0.830$ & $0.25$ \arcsec & 5.4\\
  PS z & $0.867$ & $0.803-0.935$ & $0.25$ \arcsec & 5.3\\
  PS y & $0.961$ & $0.910-1.08$ & $0.25$ \arcsec & 5.7\\
  VISTA Ks & $2.14$ & $1.93-2.37$ & $0.34$ \arcsec & 4.3\\
  IRAC 1 & $3.51$ & $3.13-3.96$ & $0.6$ \arcsec & 6.0\\
  IRAC 2 & $4.44$ &  $3.92-5.07$ & $0.6$ \arcsec & 3.0\\
  IRAC 3 & $5.63$ &  $4.90-6.51$ & $0.6$ \arcsec & 6.6\\
  IRAC 4 & $7.59$ &  $6.30-9.59$ & $0.6$ \arcsec & 5.5\\
  MIPS 1 & $23.2$ &  $19.9-30.9$ & $1.5$ \arcsec & 7.0\\
  PACS 70 & $68.9$ & $55.7-97.7$ & $1.4$ \arcsec & 8.8\\
  PACS 160 & $153.9$ & $117.8-243.6$ & $2.85$ \arcsec & 9.0\\
  
\hline\end{tabular}
\caption{Characteristics of the filters used for the study.}
\end{table*}

\subsection{Region Selection and Photometric Extraction}

 The images are all corrected for Galactic extinction using the Milky Way extinction curve of \cite{cardelli} and the colour excess (E(B-V)=0.147) from \cite{schlafly}. All images are background-subtracted using the 2D background estimator \emph{MedianBackground} from the \emph{photutils.background} library in Python.
 
The images are  degraded to the lowest resolution and pixel size of our data set which corresponds to the 160 $\mu$m image (resolution of 10.65 \arcsec). This image is processed with Scanamorphos leading to a reduction in pixel size from 6.4 \arcsec on the detector to 2.85 \arcsec in the generated image. We convolve all the other images to reach this resolution using the 2D Gaussian kernel from the Python \emph{astropy.convolution} library.
 
All the images are then re-projected with the \emph{reproject\_interp} function from the \emph{reproject} library in Python in order to fit the orientation of the  PACS images. This function also ensures that all our images have the size of the 160 $\mu$m image. At the end of the full process, all the images have the same size, orientation, pixel size and resolution. Fluxes are conserved throughout this process. 
 
The images are then divided into square boxes of four 2.85\arcsec x 2.85 \arcsec pixels resulting in square areas of 1.3 kpc$^2$ (Figure 1) which is consistent with the minimal size prescription for an unbiased SFR measurement by \cite{kennicutt}. We call each square box a 'region'.

\begin{figure*}
\centering
    \includegraphics[width=0.95\linewidth, height=0.85\paperheight]{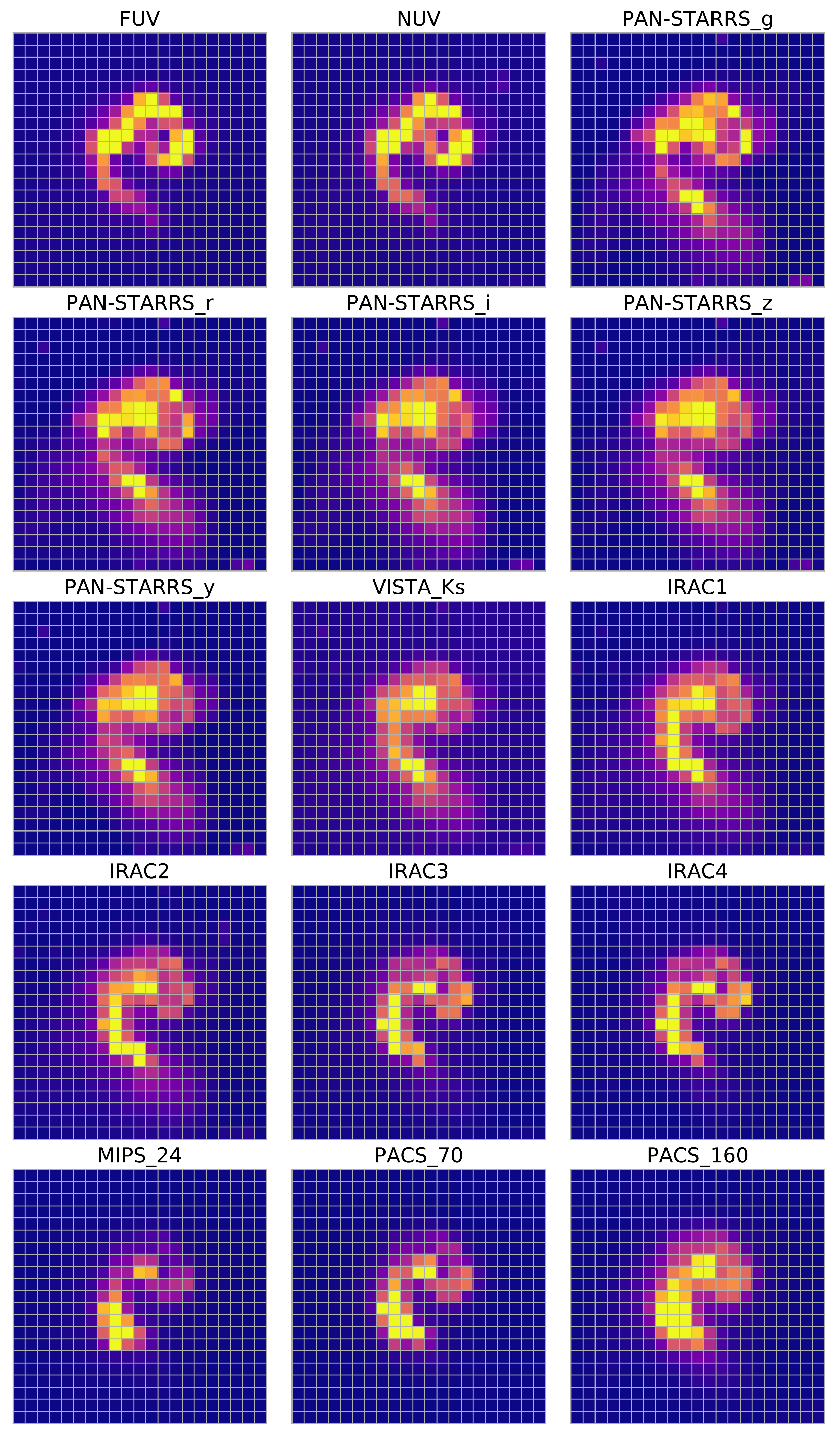}
      \caption{Images of the Antennae galaxies in the 15 different bands downgraded to the resolution of the 160 $\mu$m image.}
\end{figure*}

We want to have a good coverage of Arp 244 and a high signal to noise ratio (S/N) in all bands. We define 58 regions with a S/N higher than 3 in all bands. The minimum S/N for each band is given in Table 1. The lowest one is for the IRAC2 band. We present a map of the 58 selected regions over-imposed on the 160 $\mu$m image in Figure 2, panel (a). 

The flux  uncertainties are  estimated  by considering the background noise and  the calibration uncertainty. The background noise at each wavelength is computed  as  the mean of the flux of 66 regions located around the Antennae (see Appendix A for more details). The calibration uncertainties for the GALEX, IRAC and MIPS images are taken from \cite{zhang}: 5\% for FUV, 3\% for NUV, 10\% for IRAC, and 4\% for MIPS. The  calibration uncertainty of the PACS images is set to  5\% \citep{karl13}. The Pan-STARRS  calibration uncertainty is taken equal to 5\% instead of 1\% given by \cite{magnier}. We choose to increase the uncertainty to avoid a too large inhomogeneity with the other bands. The flux uncertainty in the Ks band is dominated by the background noise that we estimate to 20\%.

We identify the well-known broad regions of the Antennae Galaxies: NGC 4038, NGC 4039 and the Overlap Region. In Figure 2 panel (b), regions 1 to 32 and 37 to 40 correspond to NGC 4038,  regions 49, 52 to 54 and 56 to 58  to NGC 4039. The Overlap Region consists of regions 33 to 36, 41 to 48, 50, 51 and 55. We also identify the Western Arm, a part of NGC 4038 as regions 8, 15, 23, 24, 29-32, 39 and 40. 

In the following sections, we analyse  individual SEDs of  the  square regions (called 'SEDn', n being the region number)  and  the integrated SED corresponding to the sum of fluxes over the 58 regions (called integrated SED).

\begin{figure*}
\begin{subfigure}{.5\textwidth}
\centering
  \includegraphics[width=8.5cm]{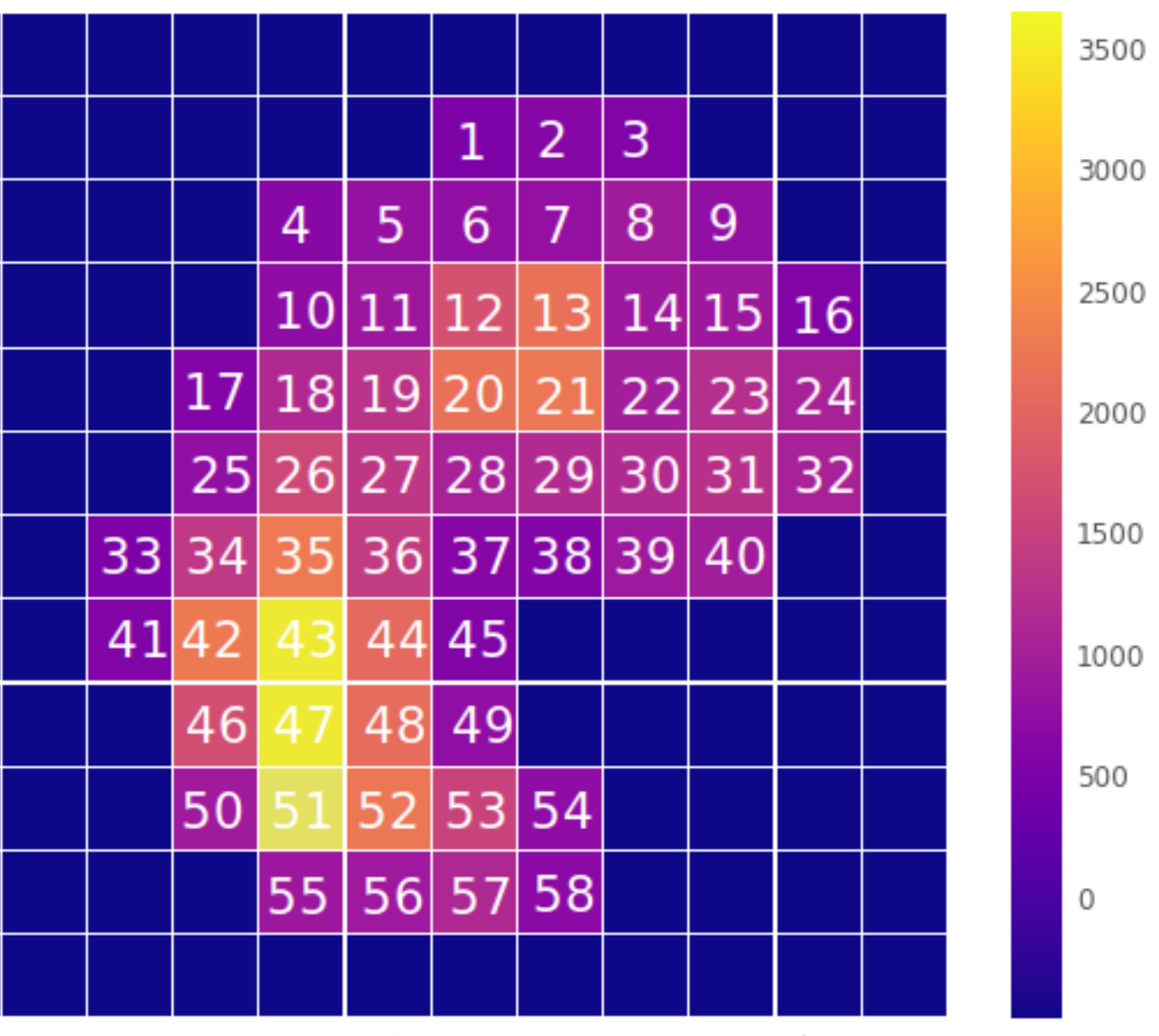}
  \caption{}
\end{subfigure}
\begin{subfigure}{.5\textwidth}
\centering
  \includegraphics[width=7.5cm, height = 7.5cm]{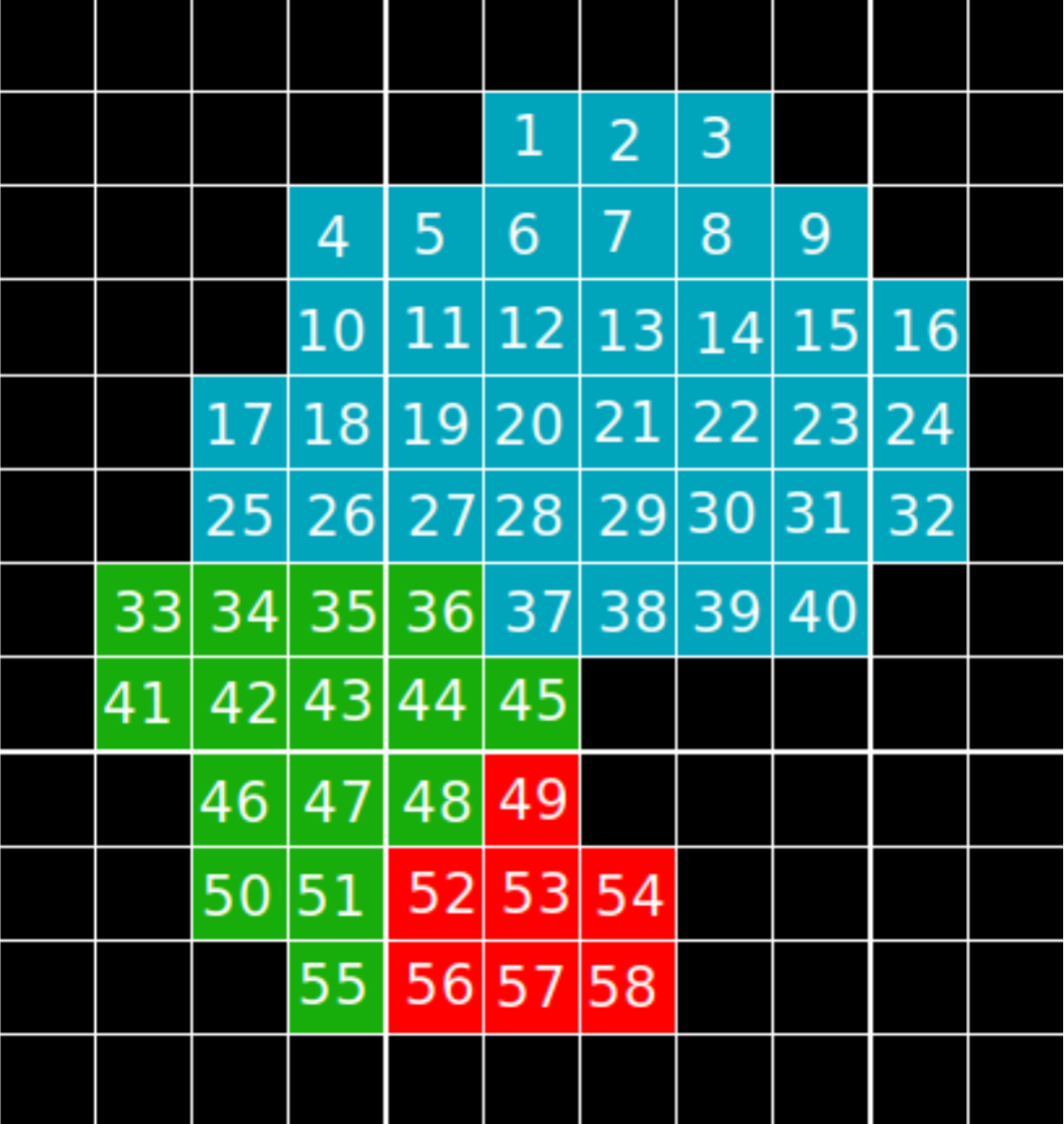}
\caption{}
\end{subfigure}
\caption{The 58 regions of the Antennae Galaxies in the 160 $\mu$m band. Each region is colour-coded with its flux in mJy (panel (a)). In panel (b), we separate Arp 244 into its three main components: regions part of NGC4038 are colour-coded in blue while regions part of NGC 4039 are colour-coded in red and regions in the Overlap are coloured in green.}
\end{figure*}

\section{Spectral energy distribution fitting}

To fit the SEDs, we use the modelling software CIGALE \footnote{https://cigale.lam.fr/version-2020.0/}. For a complete description of CIGALE and its functionalities, see \cite{boquien19}. The models are built by successively calling modules, each corresponding to a single physical component or process. CIGALE combines a UV to NIR stellar SED with a dust component emitting in the IR and conserves the energy balance between stellar dust absorption and dust re-emission. The nebular emission is added from the Lyman continuum photons produced by the stellar emission in photodissociation regions. The continuum nebular emission and the emission lines  are calculated from a grid of nebular templates \citep{villa}. The quality of the fit is assessed by the value of the $\chi^2$ and the value of the physical parameters and their corresponding uncertainties are estimated as the likelihood-weighted means and standard deviations. Below, we briefly present the modules we use and the input parameters values.

\subsection{CIGALE modules}

\subsubsection{Star formation history and stellar populations}

First we define a star formation history (SFH) and the single stellar population (SSP) model to compute the stellar emission spectrum. 
We use a  delayed exponential SFH described as:

\begin{equation}
   \mathrm{SFR} \propto t \times \exp(-t/\tau_{\mathrm{main}}),
\end{equation}
where $\tau_{\mathrm{main}}$ is the e-folding time. The module allows the addition of an optional  burst to the existing delayed exponential. The simulations of the merging of the Antennae Galaxies performed  by \cite{karl10} give an age of burst of \textasciitilde 40 Myrs after the second encounter. Thus, we define a burst of constant star formation with an age ($age_{\mathrm{burst}}$) fixed to 40 Myrs (see Sect 3.3 for more details). The amplitude of the burst is measured with the fraction of stellar mass produced in the burst, $f_{\mathrm{burst}}$. We refer to the age of the onset of star formation as $age_{\mathrm{main}}$ and we fix it to 12 Gyrs (fixing it to 13 Gyrs would not change our results). The e-folding time of the main population, $\tau_{\mathrm{main}}$, as well as $f_{\mathrm{burst}}$ are free parameters. All the input values are summarised in Table 2.

We  use the SSP  models of  \cite{bruzual} and the initial mass function (IMF) of \cite{chabrier} and we fix the metallicity to the solar value.

\subsubsection{Dust attenuation and emission}

 To compute the dust attenuation we choose a module based on the \cite{charlot} (CF00 hereafter) model. The key feature of this model is the computation of two different attenuations, one for the birth clouds (BCs) and one for the interstellar medium (ISM). In this way we are able to properly account for an age-dependent attenuation where not only the total amount of dust attenuation changes as a function of the stellar age but also the way the stellar light is extinguished, as originally introduced by CF00. We use a modified version of CF00, in which two different power law attenuation curves are used to compute the total attenuation:  

\begin{equation}
\begin{split}
    A_\uplambda^{\mathrm{BC}} & = A_\mathrm{V}^{\mathrm{BC}} (\uplambda/0.55)^{n^{\mathrm{BC}}}, \\
    A_\uplambda^{\mathrm{ISM}} & = A_\mathrm{V}^{\mathrm{ISM}} (\uplambda/0.55)^{n^{\mathrm{ISM}}},
    \end{split}
\end{equation}
where $n^{\mathrm{ISM}}$ is the slope for the attenuation curve of the ISM and $n^{\mathrm{BC}}$ the slope for the birth clouds. The slope of the attenuation curve in the birth clouds is  fixed to a reference value of -0.7 \citep{charlot,lofaro}. The power law exponent of the variation of the effective attenuation in the ISM, $n^{\mathrm{ISM}}$, is also fixed to -0.7 in the original recipe of CF00. However, we decide to treat it as a free parameter in our analysis since it has been shown to vary among galaxies e.g. \citep{buat12, chevallard, kriek, battisti17, lofaro, salim18, trayford, pantoni, boquien22}. The $V$-band attenuation in the ISM, $A_\mathrm{V}^{\mathrm{ISM}}$, is also a free parameter. The attenuation for the birth clouds, $A_\mathrm{V}^{\mathrm{BC}}$, is computed using the $\mu$ parameter:

\begin{equation}
    \mu = \frac{A_\mathrm{V}^{\mathrm{ISM}}}{A_\mathrm{V}^{\mathrm{ISM}}+A_\mathrm{V}^{\mathrm{BC}}} ,
\end{equation}

The age threshold for the separation between young stars (still attenuated by birth clouds) and older stars (only attenuated by the ISM) is 10 Myrs.

\begin{figure}
\centering
    \includegraphics[width=9 cm]{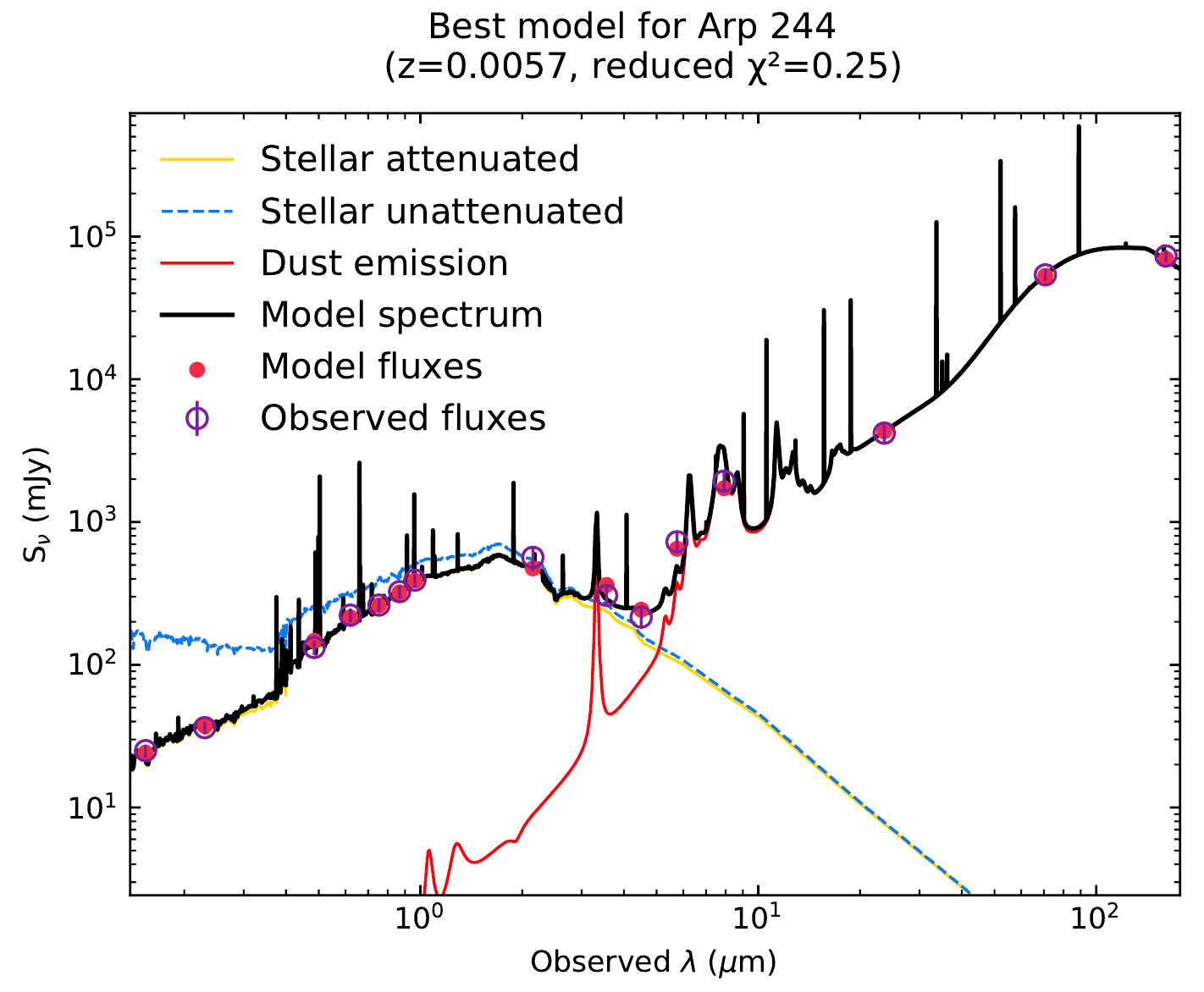}
      \caption{Integrated SED of the Antennae Galaxies. We see a high attenuation in the FUV band ($A_{\mathrm{FUV}}$ = 1.9 mag) and a moderate attenuation in the $r$-band ($A_{\mathrm{r}}$ = 0.6 mag).}
\end{figure}

To model the dust emission, we use the models from \cite{draine07} updated in \cite{draine14}. The model is built as follows: a fraction (1 - $\gamma)$ of the dust mass is heated by a starlight intensity $U_\mathrm{min}$, while the remaining fraction $\gamma$ is exposed to starlight with intensities $U_\mathrm{min}$ < U < $U_\mathrm{max}$, with a power-law distribution $dM/dU \propto$ $U^{\mathrm{-\alpha}}$. The polycyclic aromatic hydrocarbons (PAHs) abundance, $q_\mathrm{PAH}$, is also a free parameter in the models. As suggested by \cite{draine14}, we fix $U_\mathrm{max}$ to $10^{7}$ and the slope of the power law, $\alpha$, to 2. The shape of the dust emission spectrum is then determined by three free parameters: $q_\mathrm{PAH}$, $U_\mathrm{min}$, and $\gamma$. For all the input values, see Table 2.

\begin{table*}
\centering
\begin{tabular}{l r r}
\hline
\hline
  \multicolumn{1}{c}{Parameter} &
  \multicolumn{1}{c}{ Symbol} &
  \multicolumn{1}{c}{Range}
  \\
\hline
\multicolumn{1}{c}{Delayed SFH}\\
\hline
  age of the main population & $age_{\mathrm{main}}$ & 12000\\
  e-folding timescale of the delayed SFH & $\tau_{\mathrm{main}}$ & 1000-8000\\
  age of the burst & $age_{\mathrm{burst}}$ & 40\\
  burst stellar mass fraction & $f_{\mathrm{burst}}$ & 0-0.1\\
\hline
\multicolumn{1}{c}{Stellar populations synthesis}\\
\hline
initial mass function & IMF & Chabrier\\
metallicity & Z & 0.02\\
\hline
\multicolumn{1}{c}{Dust attenuation}\\
\hline
  V-band attenuation in the ISM & $A_\mathrm{V}^{\mathrm{ISM}}$ & 0.1-3.5\\
  $A_\mathrm{V}^{\mathrm{ISM}}$ / ($A_\mathrm{V}^{\mathrm{ISM}}$ + $A_\mathrm{V}^{\mathrm{BC}}$) & $\mu$ & 0.5\\
  power law slope of dust attenuation in the BCs & $n^{\mathrm{BC}}$ & 0.7\\
  power law slope of dust attenuation in the ISM & $n^{\mathrm{ISM}}$ & 0-1.2\\
\hline
\multicolumn{1}{c}{Dust emission}\\
\hline
  Mass fraction of PAHs &  $q_\mathrm{PAH}$ & 0.47-5.95\\
  Minimum radiation field & $U_\mathrm{min}$ &  1.0-15.0\\
  Power law slope of the radiation field & $\alpha$ & 2.0\\
  Fraction illuminated & $\gamma$ & 0.001-0.05\\
\hline
\hline
\end{tabular}
\caption{CIGALE modules and input parameters used for all the fits.}
\end{table*}

\subsection{Fitting the broadband SEDs}

The median of the reduced $\chi^2$ for our 58 regions is 0.65 with a minimum of 0.27 and a maximum of 2.19. Each fit is checked individually to ensure that our sample is overall well-fitted. However, the values and uncertainties of the physical parameters used in this study are estimated from their probability distribution function which account for the likelihood  of each model. We present the integrated SED of the Antennae Galaxies in Figure 3 and the SEDs of two regions in Figure 4. The infrared luminosity of Arp 244 is $8 \times 10^10 L_{\odot}$, the SFR is at 8.5 $M_\odot$yr$^{-1}$ and the stellar mass is $4.5 \times 10^{10} M_{\odot}$.

\begin{figure*}
\centering
\begin{subfigure}{.5\textwidth}
  \centering
  \includegraphics[width=8cm]{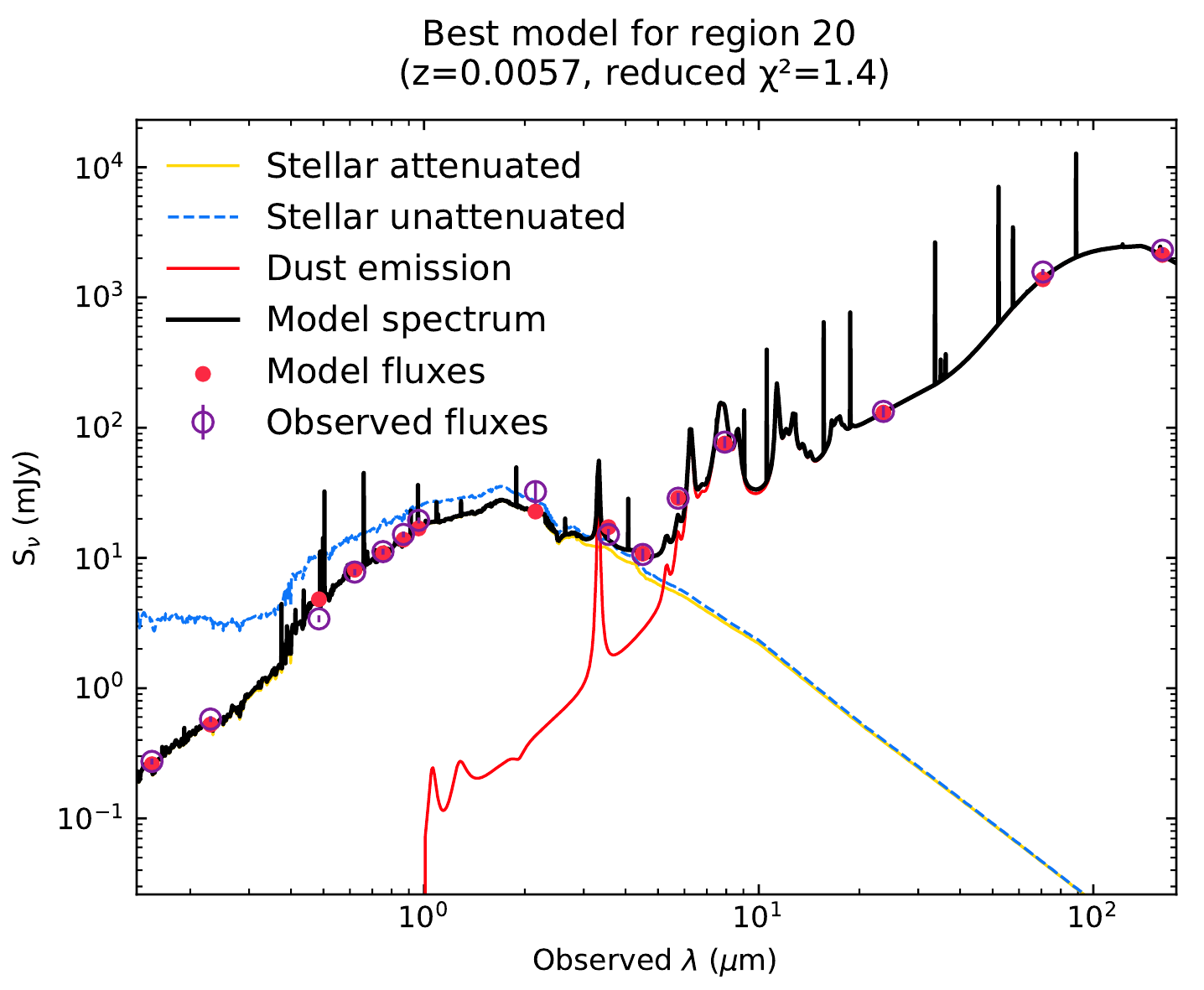}
  \label{fig:natural_weight}
\end{subfigure}%
\begin{subfigure}{.5\textwidth}
  \centering
  \includegraphics[width=8cm]{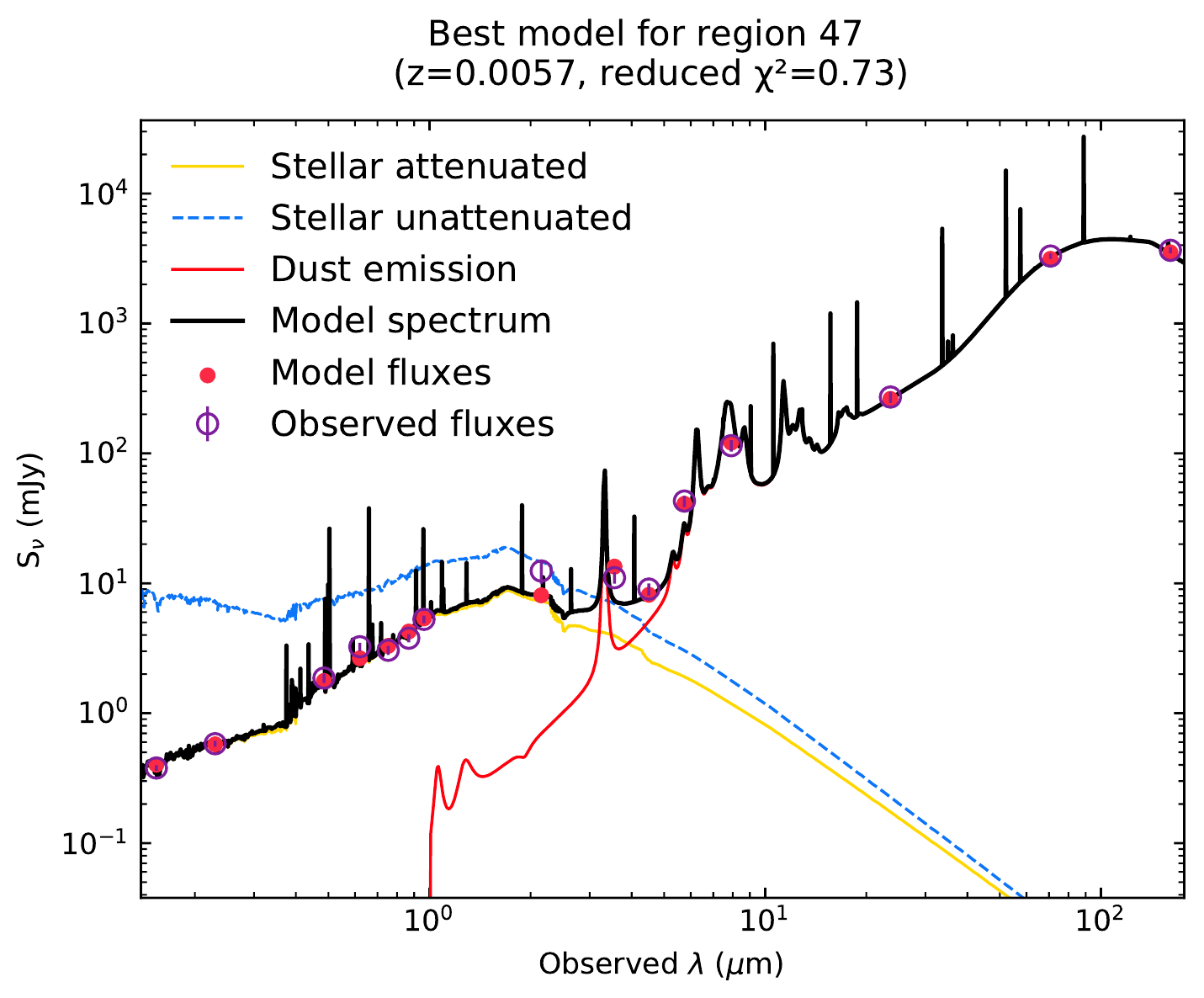}
  \label{fig:uniform_weight}
\end{subfigure}
\caption{Comparison of the SEDs from a region of the nucleus of NGC4038 (left panel) and a region part of the Overlap (right panel).}
\label{fig:weighting-comparison}
\end{figure*}

Depending on the region considered, the SEDs can exhibit completely different features. We illustrate these contrasted situations in Figure 4 with two  very different environments: a region of the nucleus of NGC 4038 (region 20) and a region which is part of the Overlap between both galaxies (region 47). Two features stand out while comparing these two SEDs: the high intrinsic emission in the UV band and the higher emission in both PACS bands for region 47 compared to region 20. As part of the nucleus of a previously spiral galaxy, region 20 is found to have a low SFR (0.18 $M_\odot$yr$^{-1}$) while region 47 has the second highest SFR of the study (0.51 $M_\odot$yr$^{-1}$). 

The Overlap Region hosts numerous star forming regions \citep{Whitmore} leading to the observed level of intrinsic UV emission in region 47. A large fraction of the UV light emitted is absorbed by dust ($A_{\mathrm{FUV}}$ = 3.4 +/- 0.2 mag) as shown in the SED and re-emitted in the infrared. In region 20, this effect still occurs but on a lower level ($A_{\mathrm{FUV}}$ = 2.6 +/- 0.2 mag). 

\subsection{Mock analysis and parameter determination}

Before discussing the estimates of the physical parameters in detail in Section 4, we want to check the validity and accuracy of our estimates through the mock analysis described below.

With CIGALE we can generate catalogues of artificial sources for which the physical parameters are known \citep{buat14,ciesla}. To build the mock catalogue CIGALE uses the best-fit model of each of the objects previously obtained through our SED-fitting procedure. The flux densities of the mock SEDs are computed by randomly picking a flux value from the normal distribution generated using as mean value the best model flux and as standard deviation the photometric error. CIGALE is then run on this artificial catalogue with the same configuration as for the first run  in order to compare the exact values of the physical parameters corresponding to the artificial SEDs to the parameters estimated by the code with the probability density function of each parameter. The analysis consists in comparing the results of the Bayesian-like analysis provided by CIGALE on this mock catalogue with the input parameters from the best models. 

We use this test to check the robustness of the output SED fitting parameters  of interest for our study and our ability to constrain them. We are mostly interested in studying the parameters linked to star formation and attenuation as the latter is critical in a dust-obscured object such as the Antennae Galaxies. Below, we describe each free parameter individually, its robustness and the need for a more in-depth discussion in the following sections.

\begin{figure*}
\centering
\begin{subfigure}{.3\linewidth}
\includegraphics[width=6.3cm]{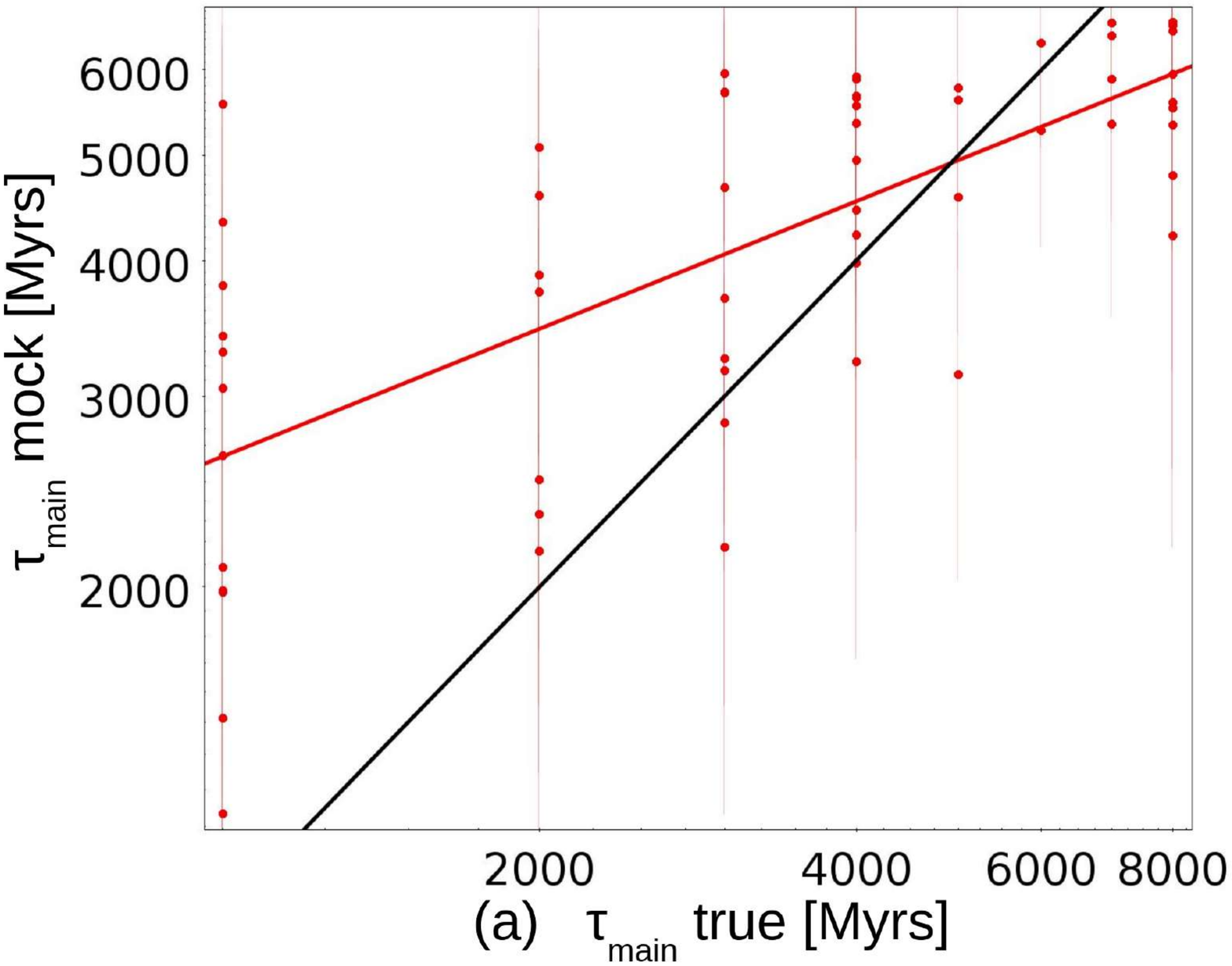}
\end{subfigure}
\begin{subfigure}{.3\linewidth}
\includegraphics[width=6.4 cm]{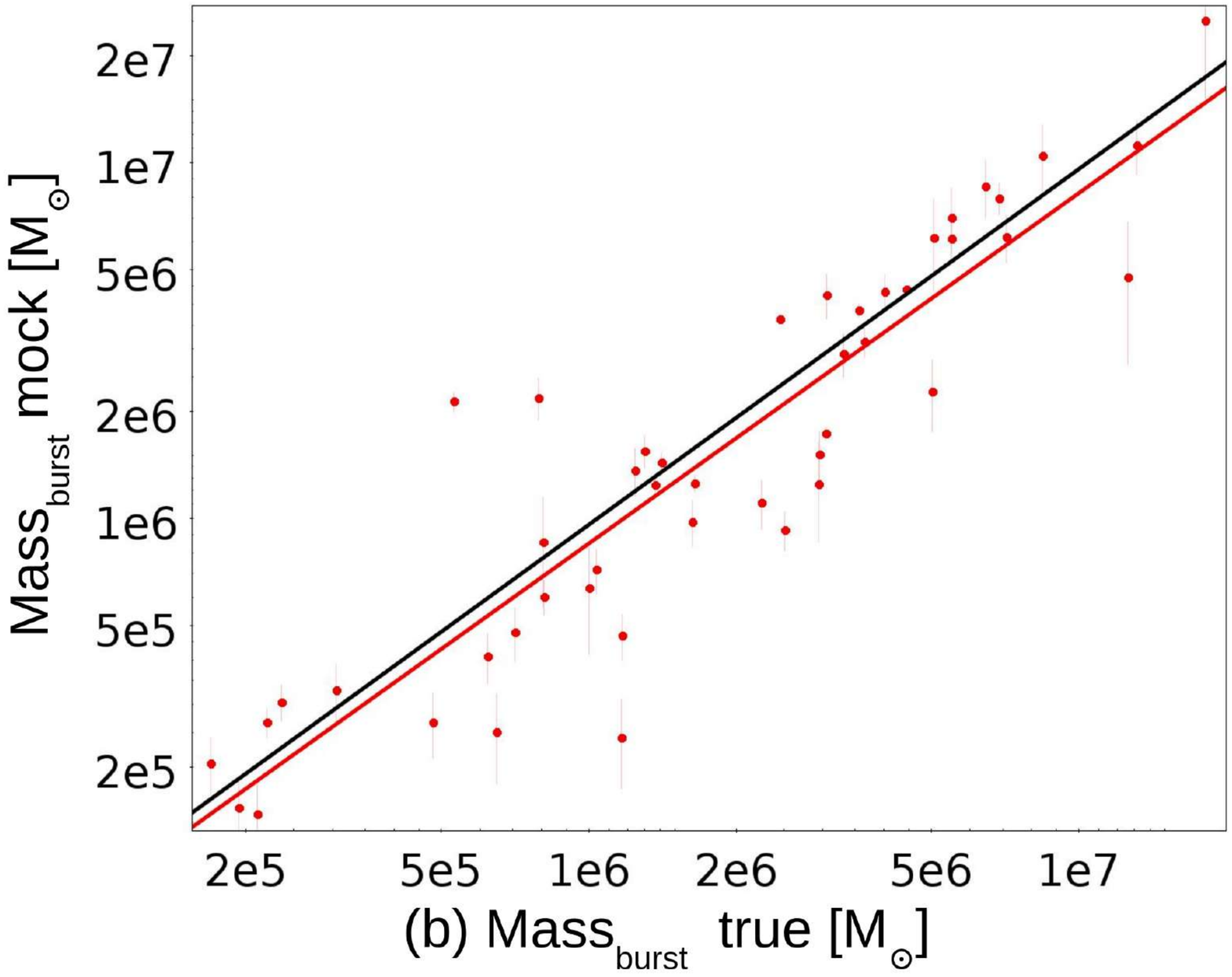}
\end{subfigure}
\begin{subfigure}{.3\linewidth}
\includegraphics[width=6.5 cm]{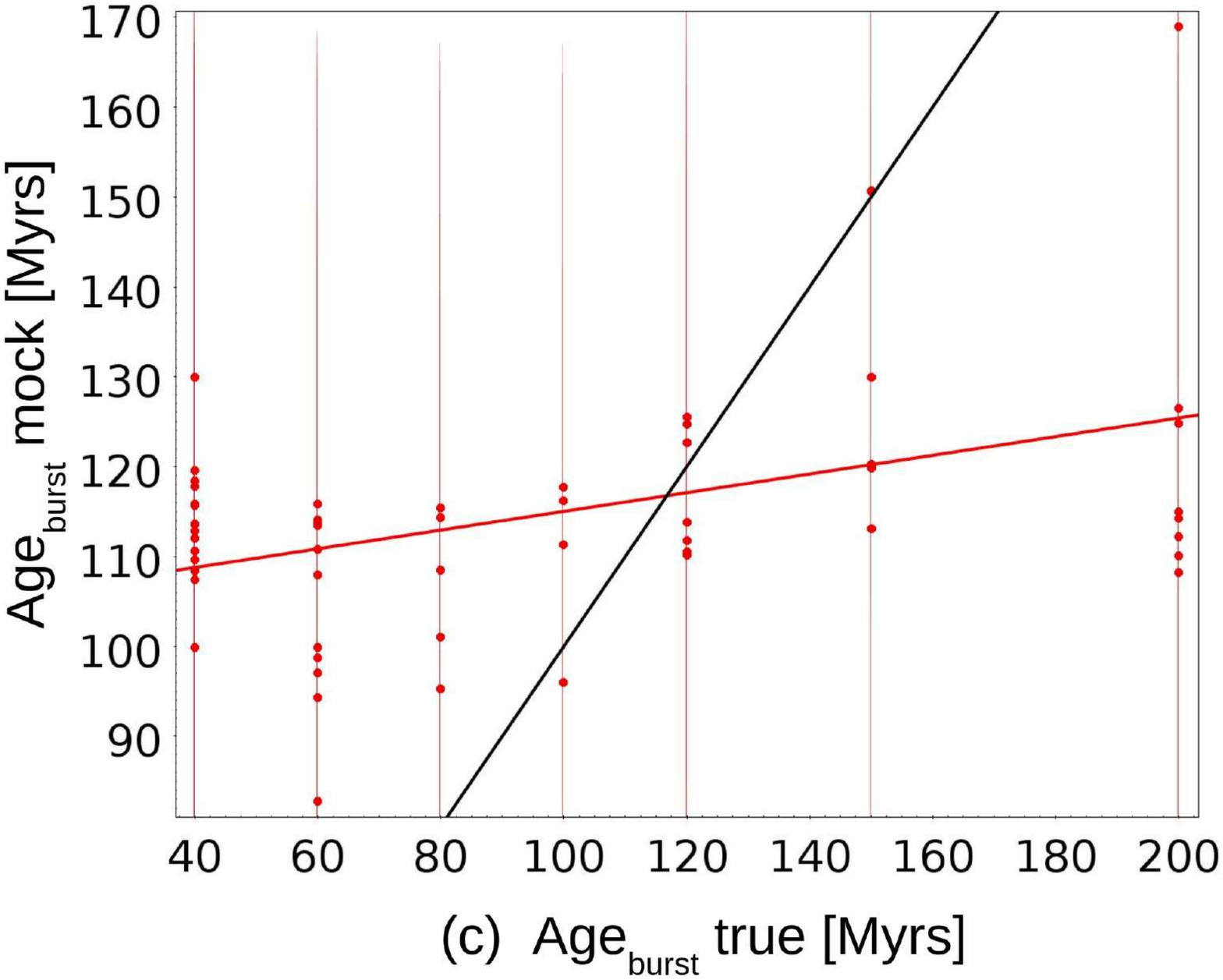}
\end{subfigure}
\begin{subfigure}{.3\linewidth}
\includegraphics[width=6.3 cm]{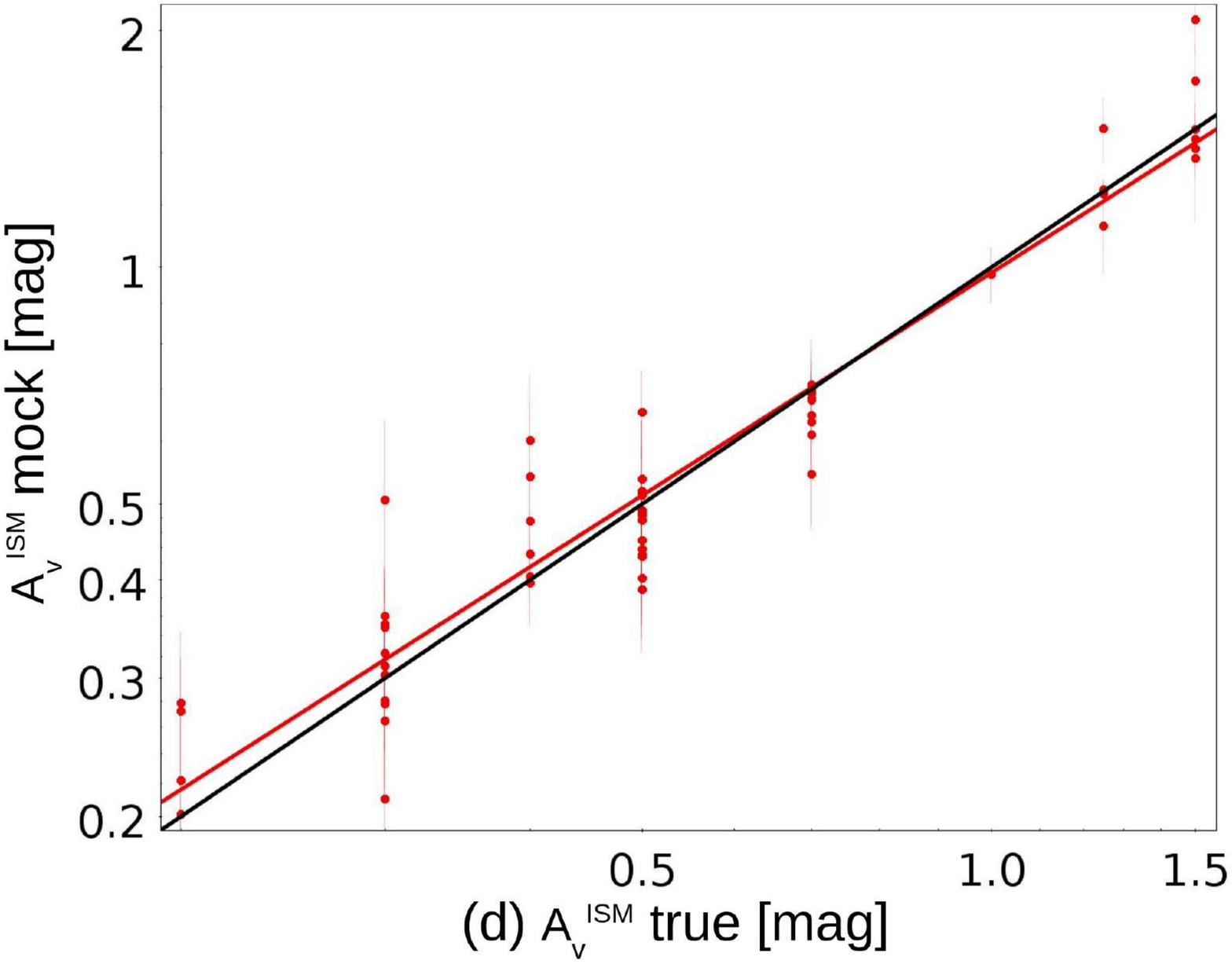}
\end{subfigure}
\begin{subfigure}{.32\linewidth}
\includegraphics[width=6.3cm]{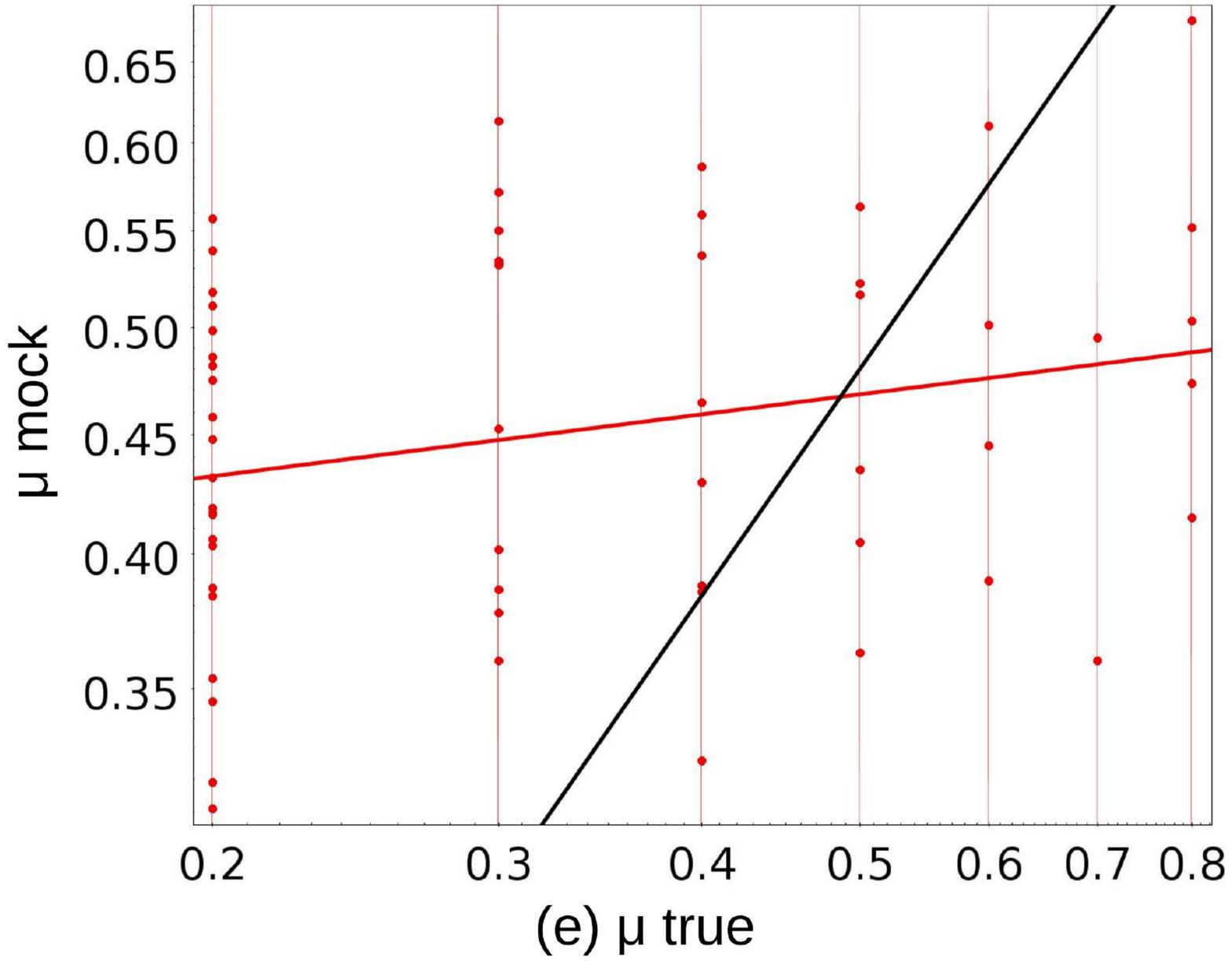}
\end{subfigure}
\begin{subfigure}{.29\linewidth}
\includegraphics[width=5.9cm, height = 4.6 cm]{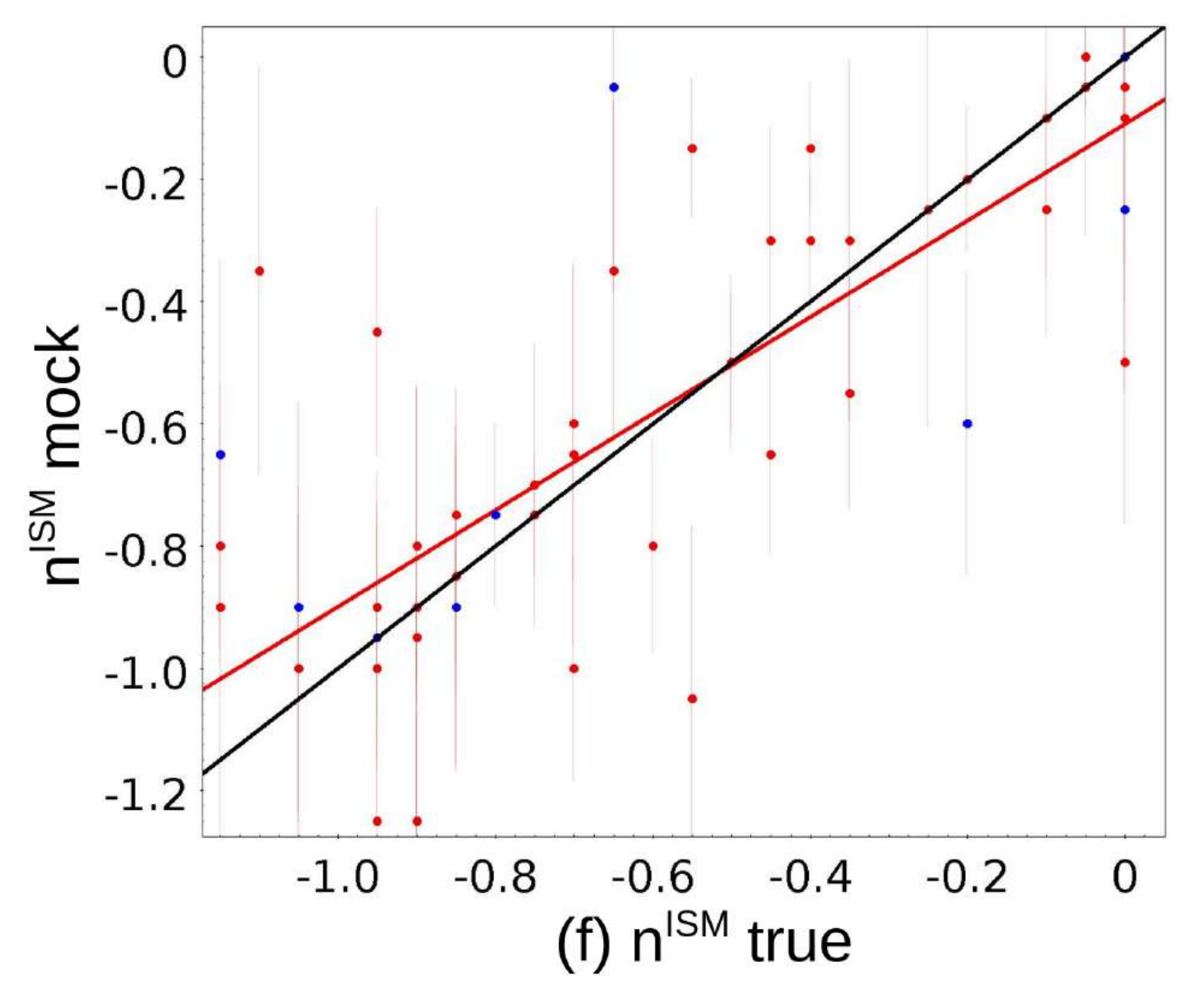}
\end{subfigure}
\begin{subfigure}{.3\linewidth}
\includegraphics[width=6.4cm]{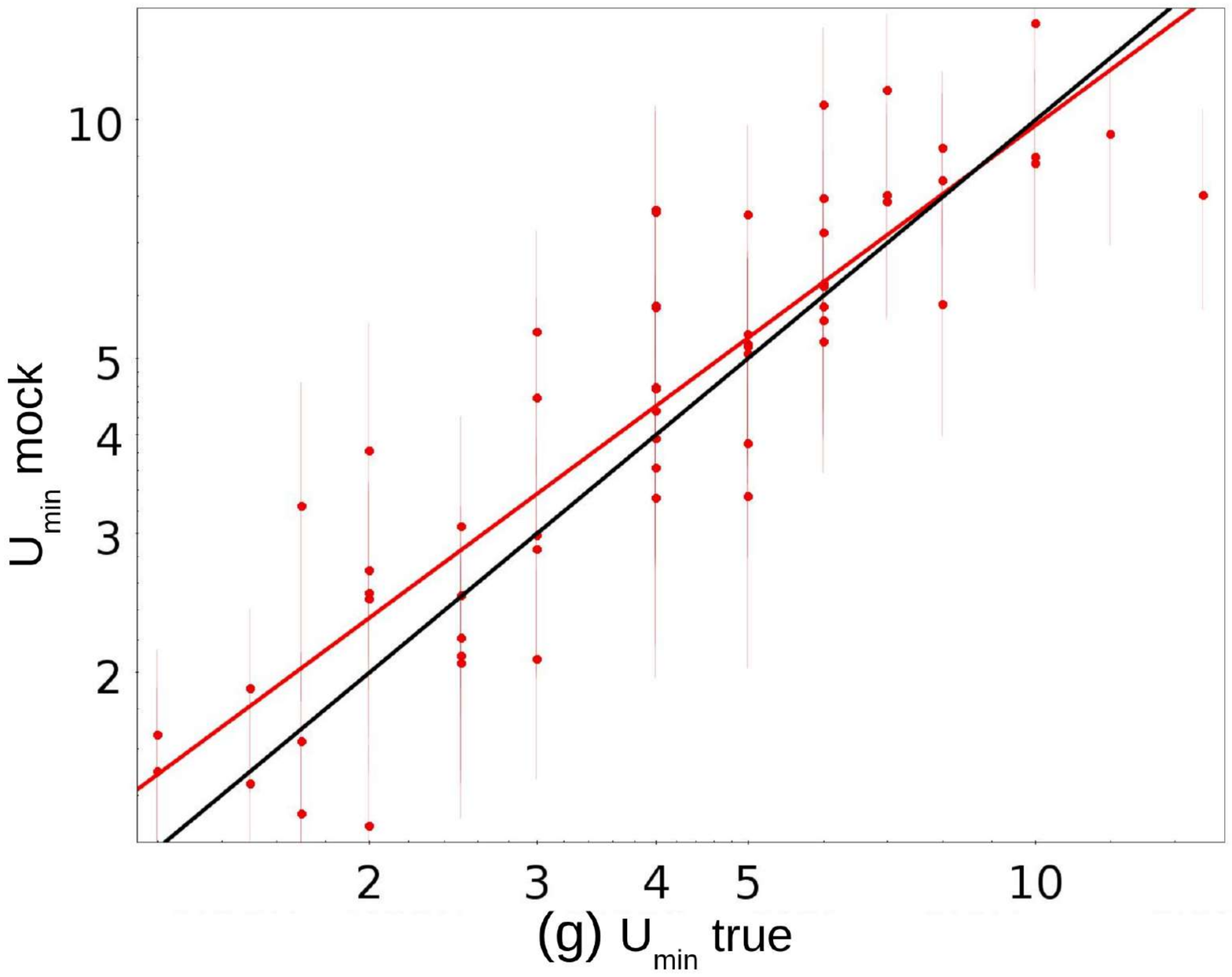}
\end{subfigure}
\begin{subfigure}{.3\linewidth}
\includegraphics[width=6.5cm]{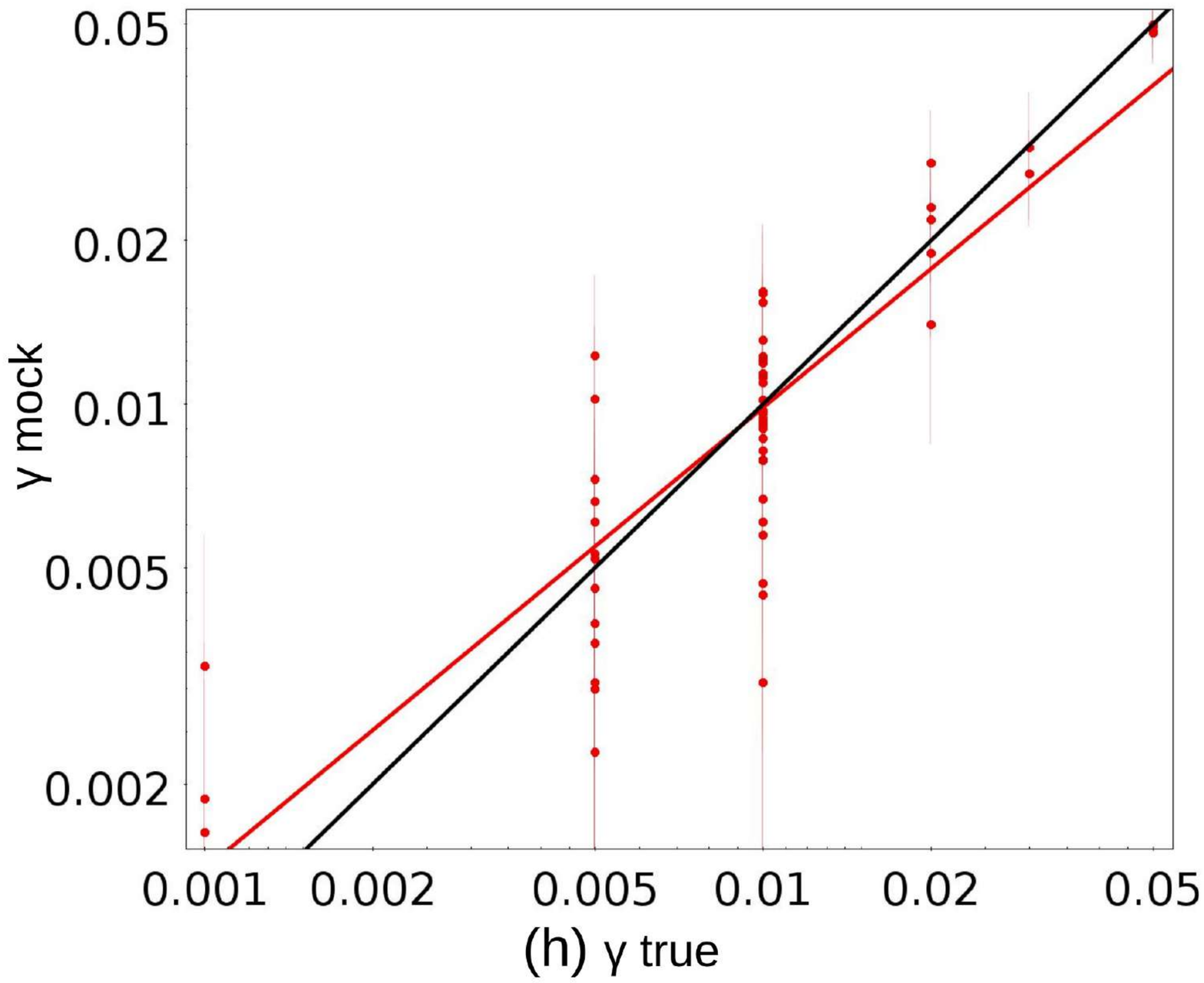}
\end{subfigure}
\begin{subfigure}{.3\linewidth}
\includegraphics[width=6.5cm]{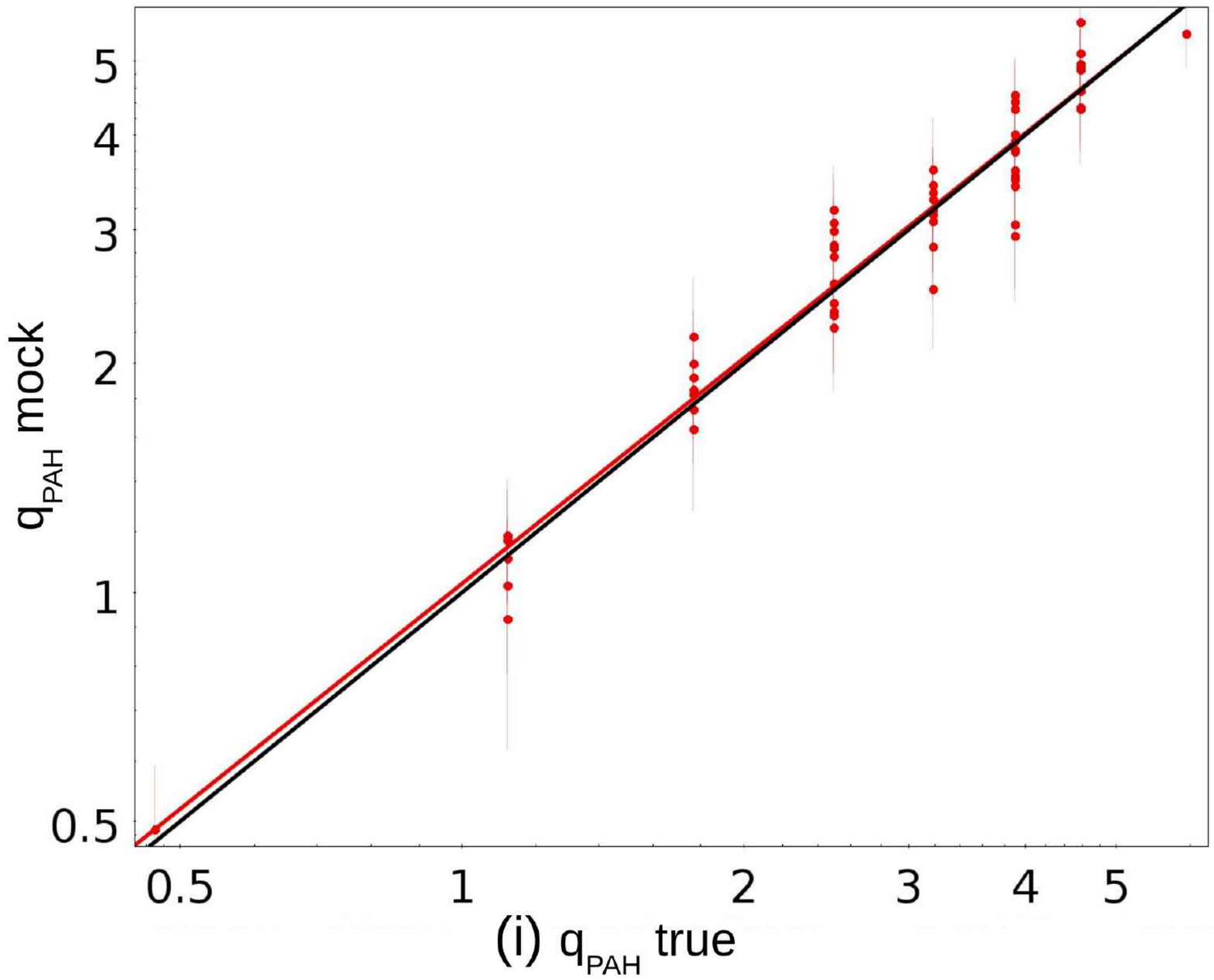}
\end{subfigure}
\caption{Results of the mock analysis. The input parameters used to build the mock catalogue are shown on the x-axis while the results of the fitting of the mock catalogues are shown on the y-axis with their associated error bars. The black solid line indicates the one-to-one relationship. The red solid line represents the linear fit of our points. The blue dots in panel f are the regions with $A_\mathrm{V}^{\mathrm{ISM}}$ < 0.3 mag.}
\label{fig:mocks}
\end{figure*}

The first parameter is $\tau_{\mathrm{main}}$, the e-folding time of the main population. As seen in Figure 5, panel (a), the comparison between the true value and its estimation on mock catalogues suggests that this parameter is not well constrained, with only a loose trend between both quantities. We decided to keep it as a free parameter to introduce  some flexibility in the fits but we will not discuss the values of this parameter.

We define the stellar mass produced during the burst, Mass$_{\mathrm{burst}}$, as the product of the Bayesian estimation of $f_{\mathrm{burst}}$ and of the total stellar mass.  $f_{\mathrm{burst}}$  is varying from 0 (no recent burst) to 0.05 (5\% of the stellar mass formed during the last 40 Myrs). Mass$_{\mathrm{burst}}$ is well-constrained (Figure 5, panel (b)) so we keep $f_{\mathrm{burst}}$ as a free parameter.

In Figure 5, panel (c), we can see that the age$_{\mathrm{burst}}$ parameter is not constrained, therefore we decided to fix it. Several ages of encounter are quoted in the literature: e.g. 6 Myrs after second pericenter passage \citep{renaud}, 40 Myrs after the second passage \citep{karl10} and various stages just before or after the second passage \citep{privon}. In our study, the most recent star formation history in traced by the FUV emission which is only sensitive to timescales of several tens of Myrs \citep{boquien14}, therefore we decided to fix age$_{\mathrm{burst}}$ to 40 Myrs.

The input parameter quantifying the attenuation in the ISM, $A_\mathrm{V}^{\mathrm{ISM}}$, is well constrained (Figure 5, panel (d)). In the following sections we will focus more on $A_\mathrm{V}$, which is the effective attenuation in the V-band. As $\mu$ is fixed (see below), $A_\mathrm{V}$ is strongly correlated to $A_\mathrm{V}^{\mathrm{ISM}}$: indeed the average ratio of $A_\mathrm{V}$ / $A_\mathrm{V}^{\mathrm{ISM}}$ is 1.1 which makes the solid constrain on $A_\mathrm{V}^{\mathrm{ISM}}$ also valid for $A_\mathrm{V}$. 

The  $\mu$ parameter which defines the effective attenuation in the birth clouds (EQ 3) is not constrained (Figure 5, panel (e)). As we do not trace the very young stars, the lack of constrain on $\mu$ was expected so we decided to fix $\mu$ to a reference value of 0.5 \citep{malek}. We checked that there is  no noticeable difference in the attenuation, SFR and the stellar mass between a run with a fixed $\mu$ and a run with a free $\mu$. On a side note, \cite{battisti20} used a free $\mu$ and fixed power law exponents in their work to account for a variable attenuation in the birth clouds.

The slope of the attenuation curve for the birth clouds was fixed to a reference value of -0.7 \citep{charlot}. The power law exponent of the variation of the effective attenuation in the ISM, $n^{\mathrm{ISM}}$, is a free parameter in our analysis and it has been shown to vary among galaxies e.g. \citep{chevallard,trayford,pantoni}. As pointed out in \cite{corre}, the slope of the attenuation cannot be accurately estimated if the amount of dust is too low. Thus, we choose to not consider regions with a value of $A_\mathrm{V}^{\mathrm{ISM}}$ under 0.3 mag in our analysis of $n^{\mathrm{ISM}}$ but to keep them for the comparison in Sect 4.3. In Figure 5 panel (f), we see that, despite the relatively high  uncertainties on the estimation of this parameter, there is a clear correlation between the true and estimated values.  

The parameters describing the dust emission, $U_\mathrm{min}$, $\gamma$ and $q_\mathrm{PAH}$, are globally constrained (Figure 5, panel (g), (h) and (i) respectively)  so we keep them as free parameters.

\section{Physical parameters estimations}

In this section, we present and discuss the spatial distribution of the parameters considered as well-constrained in the previous section. We begin with quantities related to the SFH followed by the parameters linked to  dust. We conclude this section by focusing on the SFR, the stellar mass and a comparison between  estimates of the aforementioned parameters for the whole system and the ones obtained from the sum of the 58 regions. 

\subsection{SFH parameters}

First, we focus on the analysis of the parameters of the SFH and their variation across the galaxies. Arp 244 is experiencing a merging event with a complex morphology and mass distribution. We present a pixelated map of the Antennae (Figure 6, panel (a)) where each region is colour-coded with its value of Mass$_{\mathrm{burst}}$. Two components stand out with their high values: the Overlap Region and the Western Arm (defined in Section 2.2). \cite{zhang} also find that the regions of the Western Arm exhibit a significant bursting population. This would suggest that the Western Arm was an efficient star-forming region in the past but its dusty star-forming regions have already dissipated as we are able to observe them clearly in the UV bands.

The bottom part of the Overlap Region is  forming stars actively and we also notice regions in the far right part of the Western Arm (Figure 6, panel (e)) having a SFR 1$\sigma$ above the average. We identify region 51 as the MIR 'hotspot' described by \cite{mirabel} due to its very high SFR (0.45 $M_{\odot}$ yr$^{-1}$ kpc$^{-2}$).

The nuclei of NGC 4038 (regions 20 and 21) and of NGC 4039 (regions 52 and 57) stand out with the highest stellar masses. As can be seen on Figure 6 panel (f), the Overlap Region acts as a bridge between NGC 4038 and NGC 4039 with a stellar mass comparable to that of regions surrounding both nuclei. This shows a clear disruption in the spatial distribution of the stellar mass in these galaxies.

\subsection{Attenuation and dust emission}

The distribution of the attenuation in the V-band $A_\mathrm{V}$ is presented in Figure 6, panel (b). The Overlap Region clearly stands out due to the high ratio of IR over UV emission coming from the high attenuation levels. 

As can be seen in Figure 6, panel (c), $n^{\mathrm{ISM}}$ is much smaller than -0.7 in the Overlap Region and in NCG 4038 except for its outskirts leading to flatter attenuation curves than the reference value of CF00. On the other hand, the nuclei of both progenitor galaxies have high negative values
indicating steeper attenuation curves than the reference value of CF00. We notice a tendency for regions with a high attenuation to also have a shallower  attenuation curve in the ISM (for a more in-depth discussion, see section 5).

We notice the same pattern on the maps of the SFR and the mass of the bursting population: the Overlap Region as well as the Western Arm both exhibit above average values. While the Overlap Region stands out in the attenuation map, the Western Arm has attenuation values comparable to the rest of NGC 4038. As the dusty star-forming regions in the Overlap Region have not dissipated yet, the total attenuation is much more efficient than in any part of Arp 244. 
    
Regarding the PAHs levels (Figure 6, panel (d)), we see an homogeneous spatial distribution with two notable exceptions: the nucleus of NGC 4038 and the bottom part of the Overlap Region. The latter can be explained by the fact that PAHs are destroyed by UV radiation  coming from star formation e.g. \citep{allain,boselli}, which is very intense in this part of the galaxy. Both of these features as well as the highest values of $q_\mathrm{PAH}$ in the northern part of NGC 4038 are in agreement with the distribution of PAHs levels found by Zhang et al.(2010 and references therein).

\begin{figure*}[h!]
\centering
\begin{subfigure}{.42\paperwidth}
\includegraphics[width=8.5cm, height=0.215\paperheight]{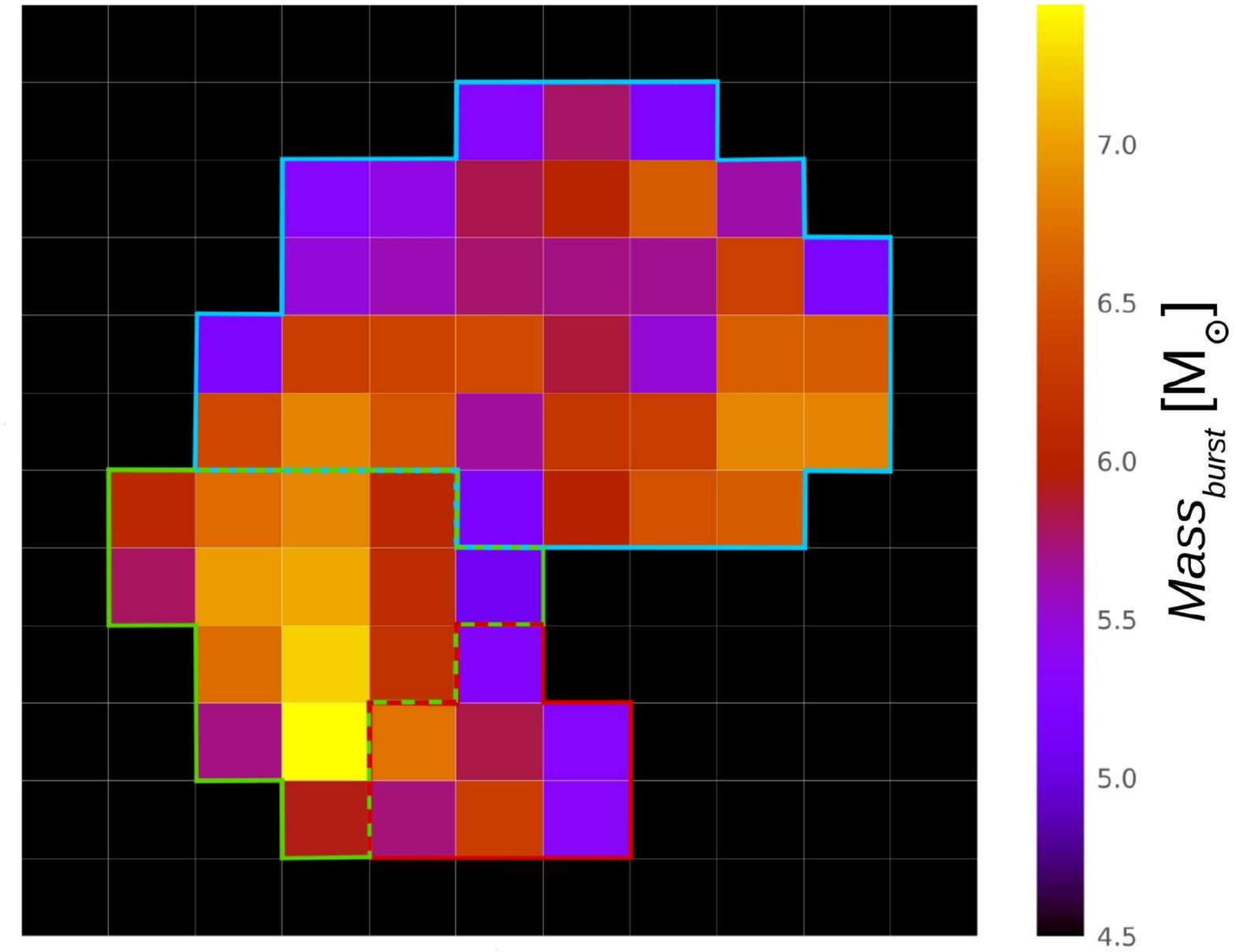}
\caption{Log Mass$_\mathrm{burst}$}\label{fig:mass_burst}
\end{subfigure}
\begin{subfigure}{.42\paperwidth}
\includegraphics[width=8.5cm, height=0.215\paperheight]{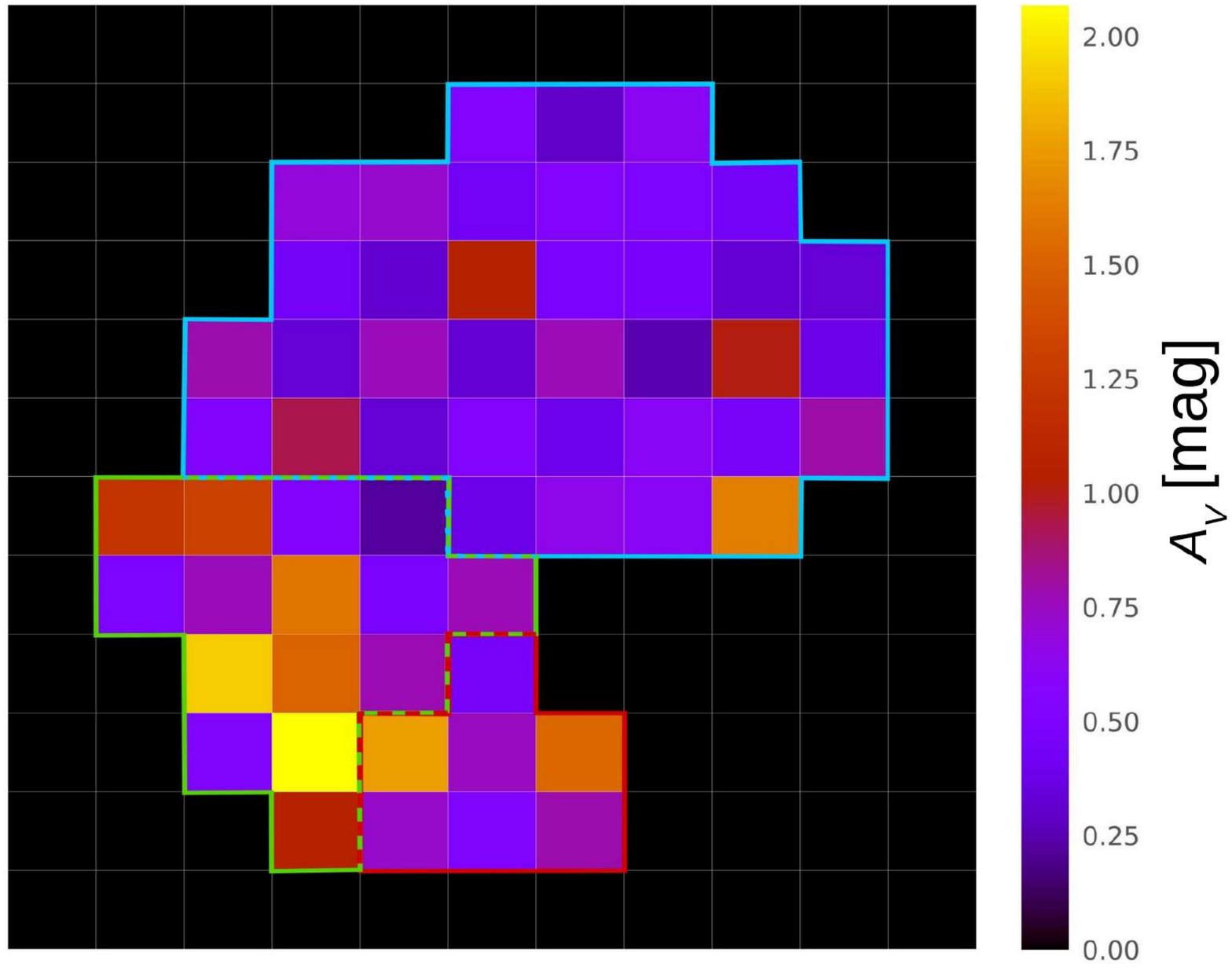}
\caption{$A_{\mathrm{V}}$}\label{fig:A_v}
\end{subfigure}
\begin{subfigure}{.42\paperwidth}
\includegraphics[width=8.5cm, height=0.215\paperheight]{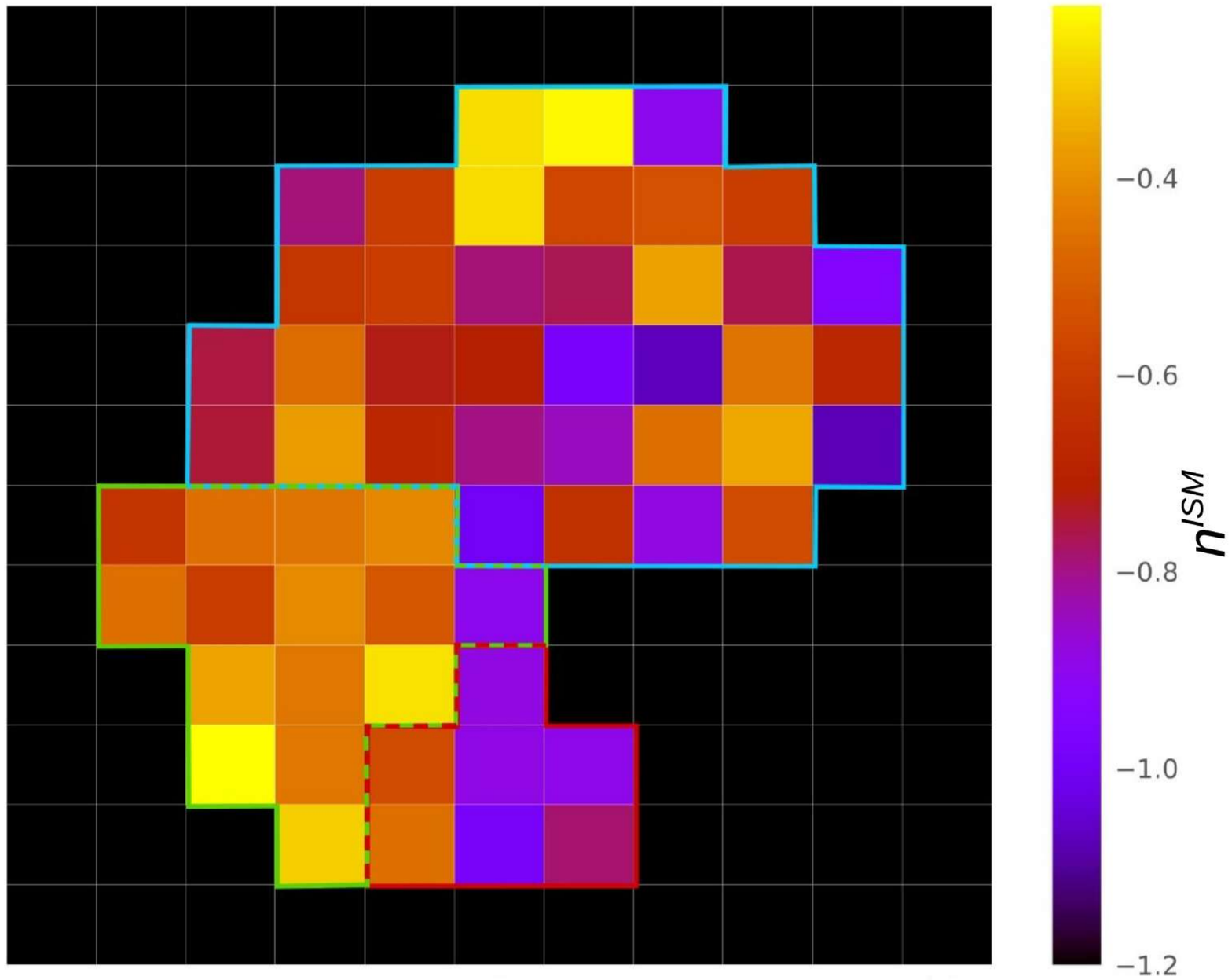}
\caption{$n^{\mathrm{ISM}}$}\label{fig:slope}
\end{subfigure}
\begin{subfigure}{.42\paperwidth}
\includegraphics[width=8.5cm, height=0.215\paperheight]{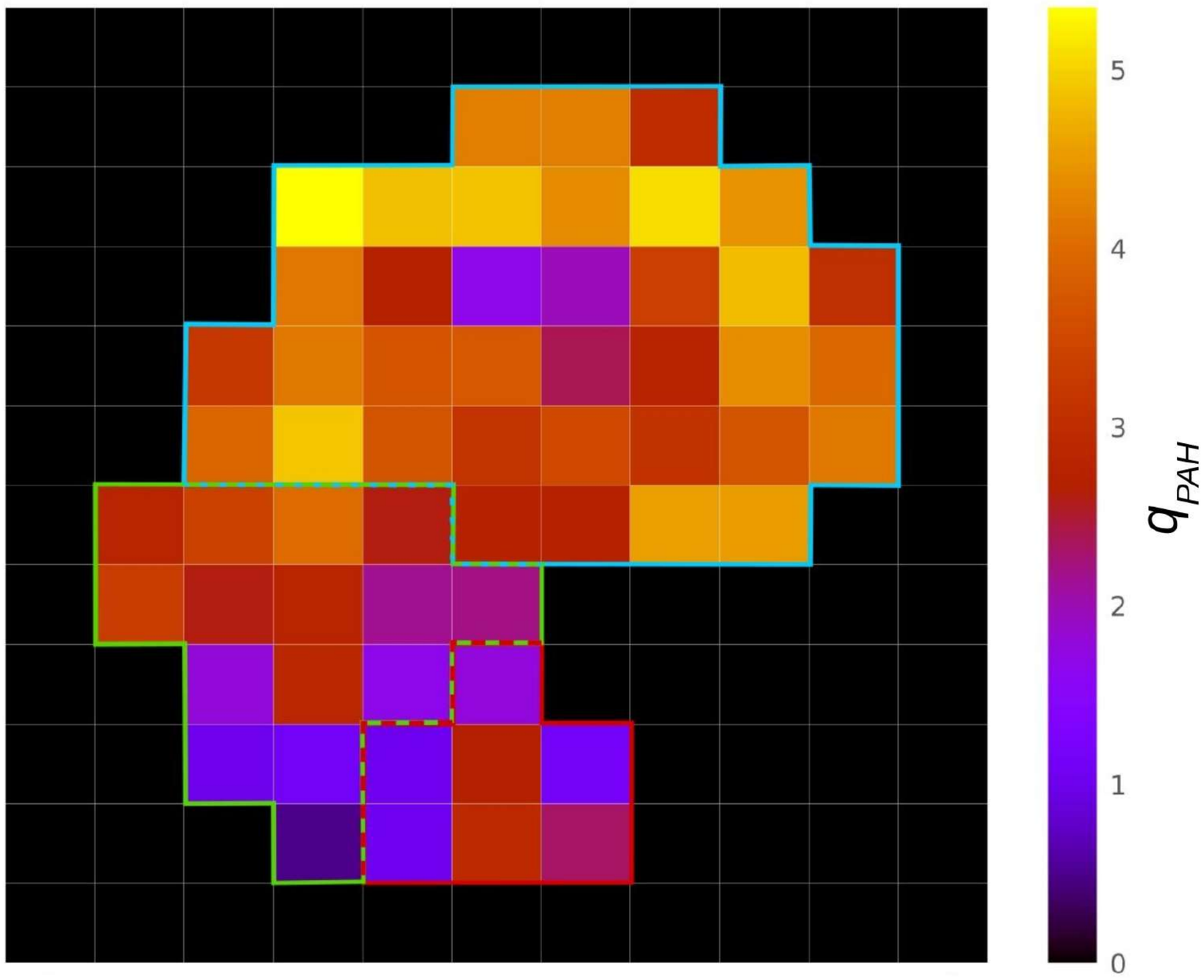}
\caption{$q_\mathrm{PAH}$}\label{fig:qpah}
\end{subfigure}
\begin{subfigure}{.42\paperwidth}
\includegraphics[width=8.5cm, height=0.215\paperheight]{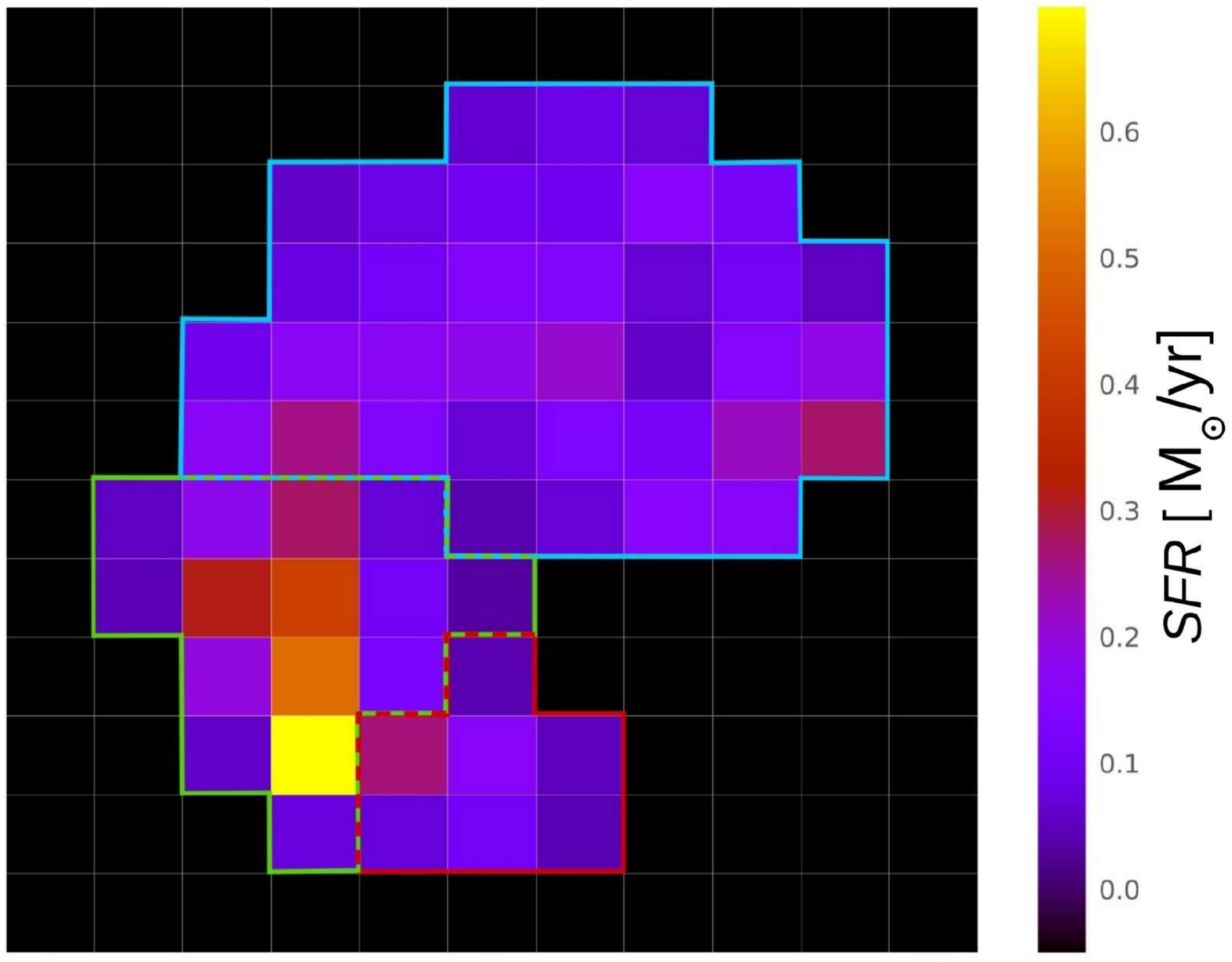}
\caption{SFR}\label{fig:sfr}
\end{subfigure}
\begin{subfigure}{.42\paperwidth}
\includegraphics[width=8.5cm, height=0.22\paperheight]{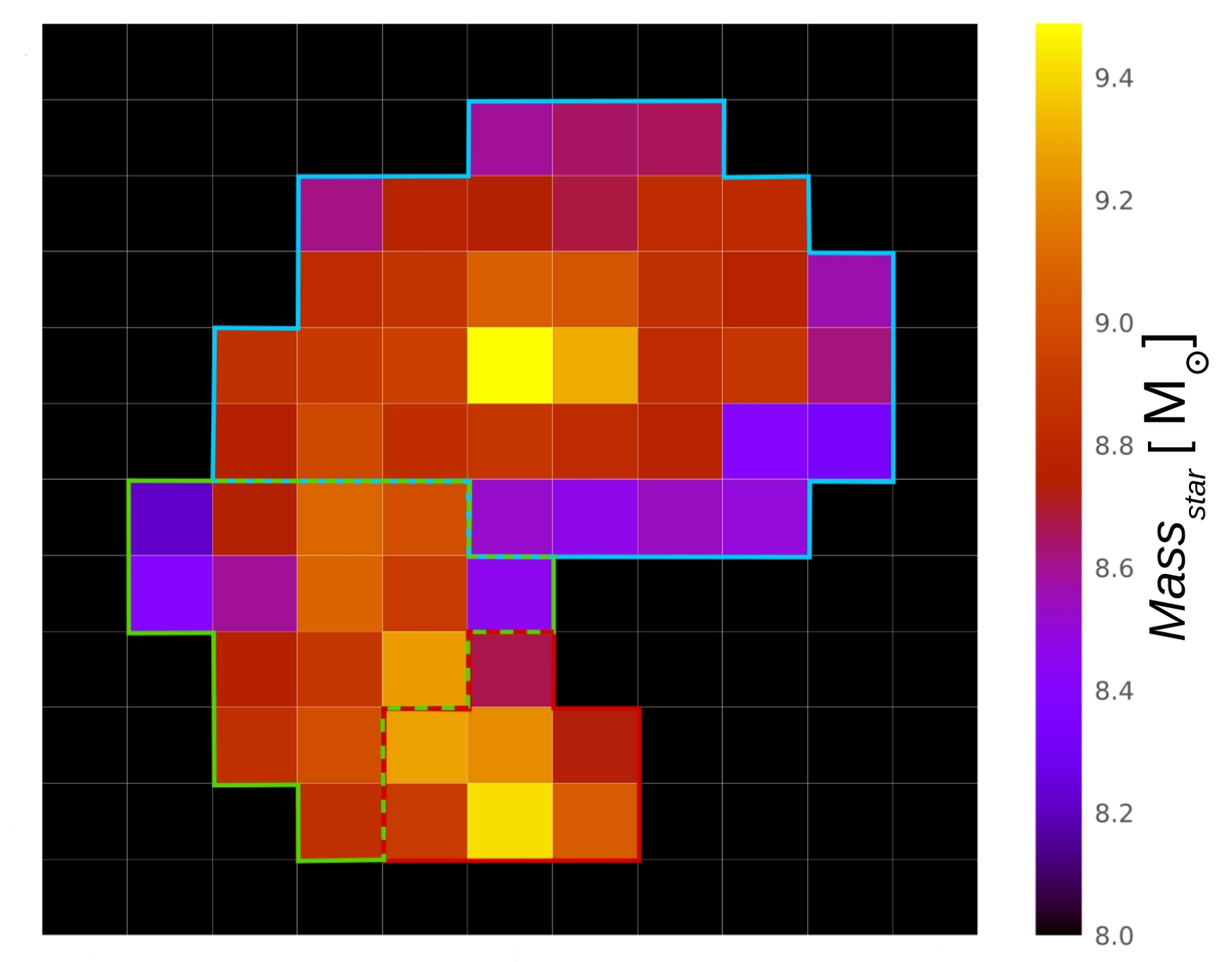}
\caption{Log Mass$_\mathrm{star}$}\label{fig:stellar_mass}
\end{subfigure}
\caption{Pixelated maps of the Antennae galaxies representing the spatial distribution of several parameters: a) mass of the bursting population, b), $V$-band attenuation, c) slope of the attenuation curve, d) $q_\mathrm{PAH}$ distribution, e) star formation rate, f) stellar mass. Coloured contours are over-plotted to separate the three components of the galaxies: blue for NGC 4038, red for NGC 4039 and green for the Overlap. Logarithmic masses have been mapped in panels (a) and (f) to allow for a better visualisation.}
\label{fig:maps}
\end{figure*}

\begin{table*}
\centering
\begin{tabular}{l r r}
\hline
\hline
  \multicolumn{1}{c}{Parameter} &
  \multicolumn{1}{c}{Mean (1$\sigma$) [min, max]} &
  \multicolumn{1}{c}{58 fluxes added (1$\sigma$)}\\
\hline
\hline
  Mass of the bursting population ($10^{7}$ $M_{\odot}$) & 17.04 ($\pm$ 4.50) [0.01, 2.76] & 12.05 ($\pm$ 2.42) \\
  V-band attenuation & 0.73 ($\pm$ 0.08) [0.23, 2.07] & 0.68 ($\pm$ 0.10)\\
  Power law slope of dust attenuation in the ISM & -0.62 ($\mp$ 0.25) [-1.07, -0.22] & -0.70 ($\mp$ 0.22)\\
  PAHs fraction & 3.16 ($\pm$ 0.46) [0.49, 5.35] & 2.86 ($\pm$ 0.58)\\
  Star formation rate ($M_{\odot}$ $yr^{-1}$) & 8.50 ($\pm$ 1.01) [0.03, 0.70] & 8.21 ($\pm$ 1.53)\\
  Stellar mass ($10^{9}$ $M_{\odot}$) & 45.8 ($\pm$ 13.0) [0.2, 3.1] & 40.5 ($\pm$ 9.5)\\
\hline
\hline
\end{tabular}
\caption{Comparison of the estimation of the physical parameters. Values of the SFR, stellar mass and mass of the bursting population of the 58 regions in the second column are added to allow for a comparison with integrated values. Thus, the minimum and maximum of these parameters correspond to the lowest and highest values out of the 58 regions.}
\end{table*}

\subsection{Comparison of SFH and dust-related estimates of 58 regions versus integrated estimates}

Here we compare the estimation of the parameters discussed above (SFR, stellar mass, mass of the bursting population, attenuation of the ISM in the $V$-band and the slope of its curve) when the values obtained for the 58 regions are added or when the whole system is fitted as a single source (Table 3).

Despite each single derived parameter spanning a wide range of values in the individual 58 regions due to the very perturbed nature of the Antennae, their mean or summed values are in agreement with those derived for the whole system. 

As the SFR in CIGALE is computed from the total infrared luminosity, we compare our value of $\log$ $(L_{IR}) = 10.89$ to the one quoted by \cite{sanders} in the IRAS sample catalogue: $\log$ $(L_{IR}) = 10.84$. Both values are close, meaning that our SFR are in agreement with the literature.

Our stellar masses estimates are only half of the value used by \cite{karl10} and \cite{lahen} to reproduce an Antennae-like system. Part of this difference may be the result of the exclusion of the tails and outer regions of the system from our computations.

Both estimates of $A_\mathrm{V}^{\mathrm{ISM}}$ and $n^{\mathrm{ISM}}$ are within the error margin. Despite hosting widely different environments, the overall attenuation and its slope are still well recovered when fitting the entire system. 

\section{Discussion}

In this section, we compare our results on  Arp 244 to several studies  of  dusty  high redshift galaxies and to the predictions of radiation transfer modelling in various galactic environments.

\subsection{Comparison with observed dusty galaxies}

\cite{lofaro}, fitted a sample of $z$ \textasciitilde 2 (ultra) luminous infrared galaxies (U)LIRGs with CIGALE. They obtained a mean power-law exponent $n^{\mathrm{ISM}}$ = -0.48 ($n^{\mathrm{BC}}$ was fixed to -0.7). The authors also fitted nearby (U)LIRGs and found that these objects are again  best fitted with a greyer attenuation curve than CF00. They concluded that local and $z$ \textasciitilde 2 (U)LIRGs are characterised by similar optical depths values and dust-clouded star forming environments which can be responsible for similar shapes in the dust attenuation curves.

At redshift $z$ \textasciitilde 2, the (U)LIRGs from \cite{pantoni} have an average specific star formation rate (sSFR) of 3.1 Gyrs$^{-1}$ while the average for non-(U)LIRGs objects is 2.2 Gyrs$^{-1}$ \citep{tasca}. Arp 244 has an average sSFR of 0.2 Gyrs$^{-1}$ which is a typical value for galaxies at redshift $z$ < 0.7 \citep{tasca}. Despite sharing spatially disconnected stellar and dust emissions, this comparison shows that some properties of $z$ \textasciitilde 2 (U)LIRGs and the Antennae Galaxies can be very different.

With the highest sSFR of the system, the Overlap regions are those with physical properties most similar to those of high-$z$ galaxies. We find that they are best fitted with greyer attenuation curves (with $n^{\mathrm{ISM}}$ between -0.22 and -0.63)  than the reference value of CF00 (-0.7). However, the value of $n^{\mathrm{ISM}}$ for Arp 244 as a whole (-0.62) is closer to that of CF00 (typical value for local galaxies) than to the one of LF17 (-0.48). This suggests that only the Overlap Region (average $n^{\mathrm{ISM}}$=-0.45) exhibits an attenuation curve exponent comparable to $z$ \textasciitilde 2 (U)LIRGs while the entire system (Arp 244) does not.   

Galaxies at $z$ \textasciitilde 2 observed with ALMA have generally disturbed morphologies with a clumpy dust emission detected at millimetric wavelengths e.g.\citep{dunlop,elbaz} significantly different and spatially dissociated from the one of the stellar populations as derived from the optical bands. We observe the same phenomenon in the Antennae Galaxies as shown in Figure 1. SED fitting of these high-$z$ galaxies suggest flatter attenuation curves than the one of CF00 ($n^{\mathrm{ISM}}$=-0.7).

\begin{figure}
\centering
    \includegraphics[width=9 cm]{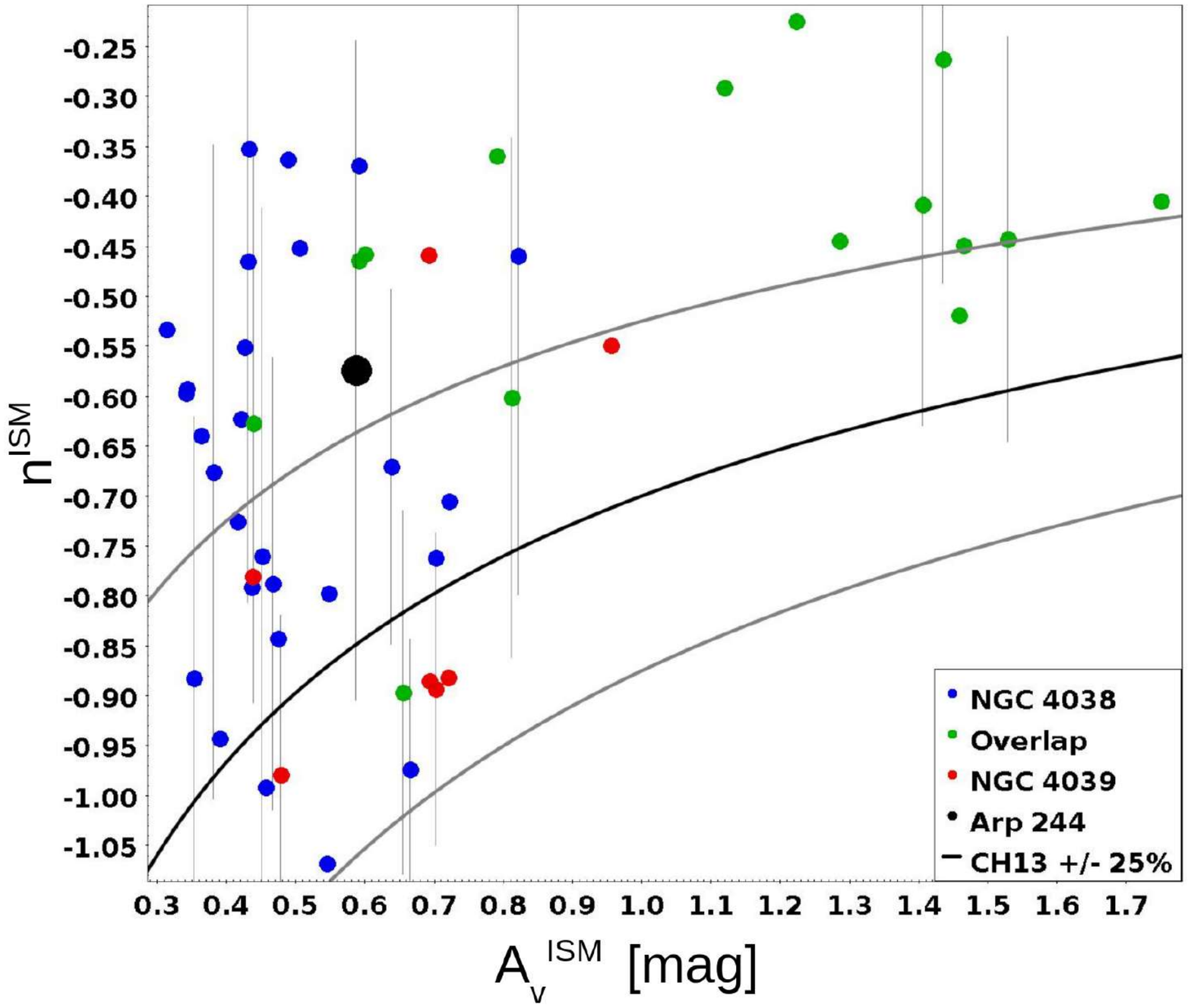}
      \caption{Slope of the attenuation curve $n^{\mathrm{ISM}}$ plotted against the optical dust attenuation of the ISM, $A_\mathrm{V}^{\mathrm{ISM}}$. Blue, green and red dots represent the 58 regions and the black dot the entire galaxy. Only one out of three error bars is plotted to keep the plot readable. The black curve represents the best fit with their associated error (grey curves, defined at 25\% by CH13).}
\end{figure}

\cite{buat19} used SED fitting on seventeen $z$ \textasciitilde 2 sources. Six galaxies are best fitted with the standard value of CF00, seven with a Calzetti et al. (2000, C00 hereafter) recipe ($n^{\mathrm{ISM}}$= -0.7) and the remaining four with a LF17 recipe. They also found that shallow attenuation curves are better suited to describe SEDs of galaxies with a dust distribution much less extended than the light distribution coming from stars. 

\cite{pantoni} performed SED fitting on 11 starbursts objects at redshift $z$ \textasciitilde 2 of the GOODS-S field. Nine objects are classified as ULIRGs, the remaining two as LIRGs. Using CIGALE, they found that 8 out of 11 of their objects are best fitted with a shallower attenuation curve than the standard assumption of $n^{\mathrm{ISM}}$ = -0.7 (CF00). They obtain a median  attenuation curve exponent of $n^{\mathrm{ISM}}$ = -0.5. which is consistent with the average value we find for the Overlap Region of the Antennae ($n^{\mathrm{ISM}}$ = -0.45).

Following these comparisons, we observe that high-$z$ galaxies exhibit overall flatter attenuation curves than CF00 while Arp 244 has an attenuation curve close to CF00. The Overlap Region is the only part of the system whose attenuation curves resemble the ones found for high redshift objects. 

\subsection{Comparison with radiative transfer models}

Radiative transfer modelling has  been extensively used to study variations of the attenuation curve in both resolved and unresolved galaxies. These studies can be broadly classified in two categories: the first one with dust / stars interactions calculated with several stellar components distributed in bulge and disks (e.g. \cite{silva}; \cite{tuffs}; \cite{pierini}) and the second one relies on hydrodynamical simulations with a post-process radiative transfer analysis (e.g. \cite{saftly}; \cite{camps16}; \cite{feldmann}; \cite{rodriguez}; \cite{roebuck}; \cite{trayford}).  Both predict that the effective attenuation curve depends on the amount of dust attenuation . 

\begin{figure*}
\begin{subfigure}{.5\textwidth}
\centering
  \includegraphics[width=9.4 cm]{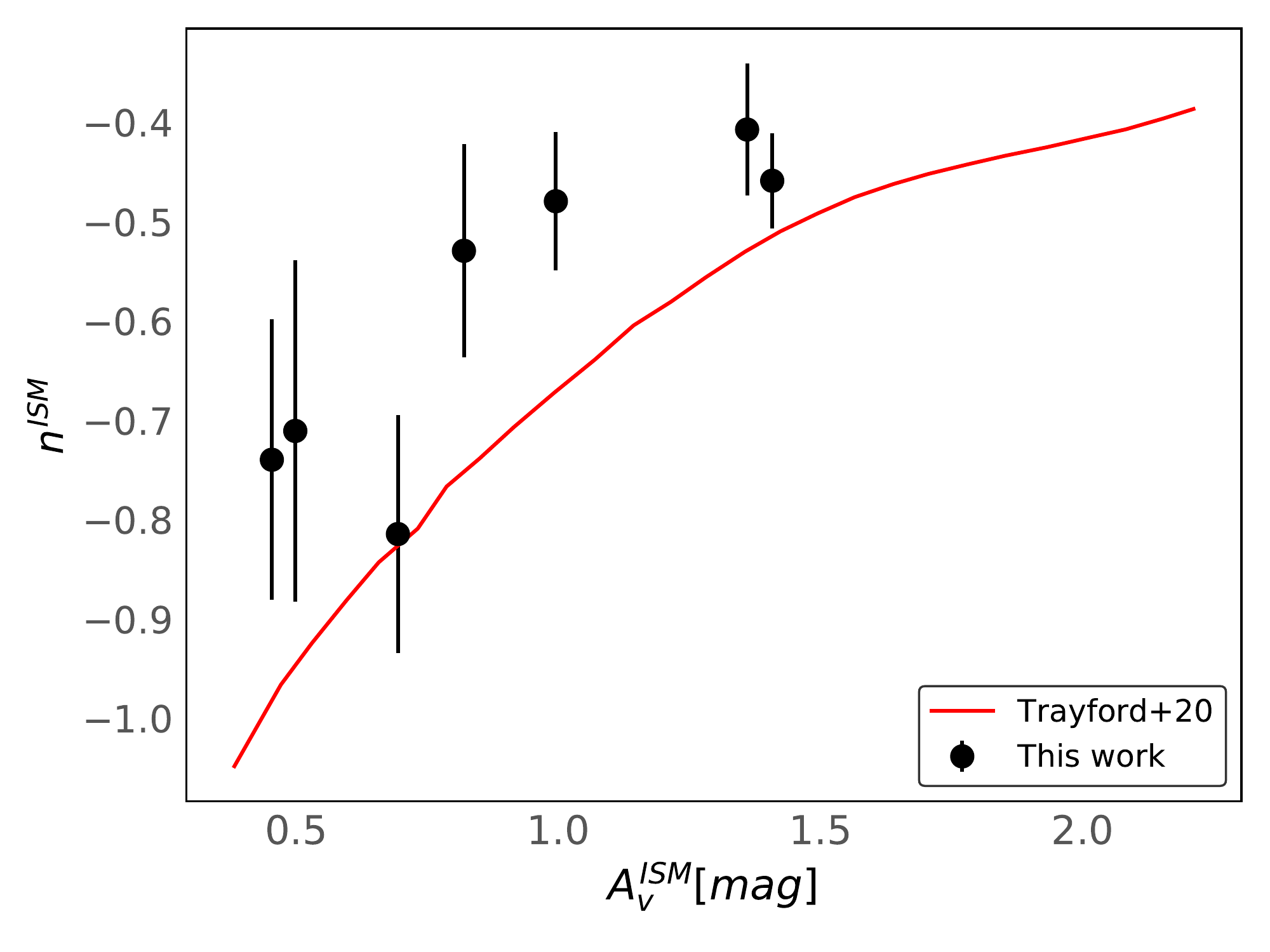}
  \caption{}
\end{subfigure}
\begin{subfigure}{.5\textwidth}
\centering
  \includegraphics[width=9.4 cm]{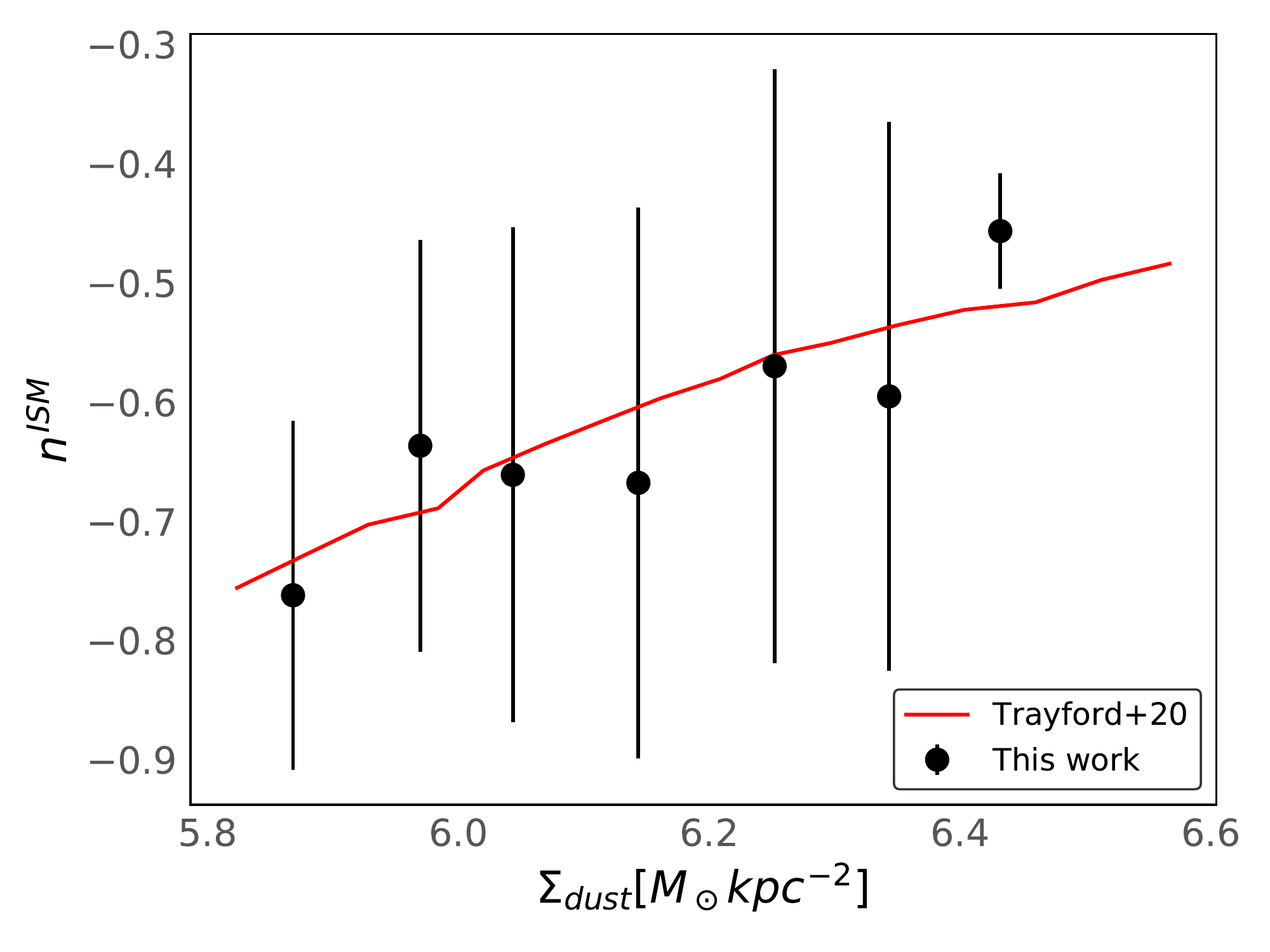}
  \caption{}
\end{subfigure}
\caption{Relation between $n^{\mathrm{ISM}}$ and dust-associated quantities. The slope of the attenuation curve $n^{\mathrm{ISM}}$ is plotted against the optical dust attenuation of the ISM, $A_\mathrm{V}^{\mathrm{ISM}}$ in panel (a) and against the dust surface density $\Sigma_{\mathrm{dust}}$, in panel (b). The black dots show the mean for each bin (0.1 separation) with its associated error bar. The red curve represents the best fit from \cite{trayford}. The mean was only computed if there was more than two objects in the selected bin.}
\end{figure*}

\subsubsection{Radiative transfer models of low redshift, resolved galaxies}

Using several radiative transfer models for galaxies, \cite{chevallard}  proposed a  relation (hereafter CH13) linking the slope of the attenuation curve and the optical dust attenuation of the ISM. In Figure 7, we plot the relation between the V-band attenuation and the exponent of the attenuation curve in the ISM for our 58 regions as well as the CH13 relation. 
 
Overall the individual regions of Arp 244 follow a similar trend between $n^{\mathrm{ISM}}$ and $A_\mathrm{V}^{\mathrm{ISM}}$ as the relation proposed by CH13. Our measured $n^{\mathrm{ISM}}$ are found higher than the model predictions inducing flatter curves for a given $A_\mathrm{V}^{\mathrm{ISM}}$ but the large uncertainties on both measurements and model predictions make them at least marginally consistent. Regions associated to NGC 4039 follow the relation more closely than those of NGC 4038, which are scattered in Figure 7. Regions from the Overlap have, on average, a lower $n^{\mathrm{ISM}}$ with respect to the one given by the CH13 relation. 

CH13 found that the main drivers of changes in the shape of their optical attenuation curves are the geometry of the dust and orientation of the galaxy. However, they use models of symmetrical, non-perturbed galaxies. While the orientation of NGC 4038 is clearly face-on, it is more difficult to assess for the Overlap Region and NGC 4039 as their morphology is very disturbed. These disturbed morphologies and complex dust geometry in the Overlap Region may be the cause for these regions falling outside of the CH13 error margin. 

\subsubsection{Radiative transfer processing of galaxy formation simulations}

In order to simplify the radiative transfer modelling, semi-analytic models of galaxy formation (e.g. \cite{fontanot}; \cite{gonzalez}) only provide the basic morphological components of galaxies (bulge and disc) which are in turn modelled using idealised geometries . Thus, it is highly likely that they miss at least part of the influence of multi-scale clumping and inhomogenity in the galaxy structure. 
Hydrodynamical simulations, however, are able to produce structures more representative of real galaxies, like clumps and inhomogeneities. Targeted hydrodynamical simulations (e.g. \cite{saftly}; \cite{feldmann}) as well as the more recent hydrodynamical simulations of cosmological volumes (e.g. \cite{camps16}; \cite{trayford}; \cite{rodriguez}) allowed more representative attenuation properties to be produced and compared with observations.

Recent works of \cite{trayford} and \cite{roebuck} give prescriptions on the attenuation curves of high redshift galaxies which can be compared to our measurements. Using a sample of \textasciitilde 100 000 simulated galaxies up to  $z$ \textasciitilde 2 from the EAGLE simulations (\cite{schaye}; \cite{crain}), \cite{trayford} calculated the dust attenuation using the 3D radiative transfer code, SKIRT (\cite{baes03}; \citep{baes11}; \cite{camps15}). They found the same trend as \cite{chevallard}: the slope of the attenuation curve $n^{\mathrm{ISM}}$ becomes smaller than -0.7 as the optical dust attenuation $A_\mathrm{V}^{\mathrm{ISM}}$ increases. In Figure 8 panel (a), we present our results using bins with a 0.1 mag separation: the slope of the attenuation curve $n^{\mathrm{ISM}}$ becomes smaller than -0.7 with increasing optical dust attenuation $A_\mathrm{V}^{\mathrm{ISM}}$. As in Figure 7, we obtain higher $n^{\mathrm{ISM}}$ than the model predictions for a given bin of $A_\mathrm{V}^{\mathrm{ISM}}$.

\cite{trayford} also analysed the possible dependence of the slope of the attenuation curve on the dust surface density. They find that the attenuation curve becomes greyer as the dust surface density increases. Our data for $\Sigma_{\mathrm{dust}}$ only ranges from 5.8 to 6.5 $M_{\odot}$ kpc$^{-2}$, divided into bins separated by 0.1 $\Sigma_{\mathrm{dust}}$), but the trend is very similar as shown in Figure 8, panel (b).  We also find that regions with $\Sigma_{\mathrm{dust}}$ > 6 $M_{\odot}$ kpc$^{-2}$ need a greyer attenuation than the reference value of CF00. The nucleus regions ($\Sigma_{\mathrm{dust}}$ bins 6.1 to 6.2 and 6.3 to 6.4 for NGC 4039 and NGC 4038 respectively) correspond to lower values with respect to the model. All the other regions with $\Sigma_{\mathrm{dust}}$ > 6 $M_{\odot}$ kpc$^{-2}$ are either inside or adjacent to the Overlap Region. As we find the same trend as \cite{trayford} between $n^{\mathrm{ISM}}$ and $\Sigma_{\mathrm{dust}}$ but a quite a different one between $n^{\mathrm{ISM}}$ and $A_\mathrm{V}^{\mathrm{ISM}}$ for low values of $A_\mathrm{V}^{\mathrm{ISM}}$, we conclude that the relation between $\Sigma_{\mathrm{dust}}$ and  $A_\mathrm{V}^{\mathrm{ISM}}$ is different for the Antennae Galaxies compared to the galaxies from the EAGLE simulations.

\cite{roebuck} used the model library generated by the gadget-2 cosmological N-body/smoothed particle hydrodynamics (SPH) code \citep{springel}. The authors processed these models using the sunrise (\cite{jonsson06}; \cite{jonsson10}) radiative transfer suite to generate panchromatic SEDs at 10-100 Myr increments. One feature of interest in \cite{roebuck} is the comparison of the effective attenuation curves of two major merger simulated objects at redshift $z$ \textasciitilde 3 to several attenuation curve recipes. The two most interesting ones for our study are LF17 and C00: the authors plotted different attenuation curves for different epochs of the merging. The authors show that the more the merging process advances, the steeper the attenuation curve becomes. Up to the coalescence stage, the attenuation curves of both of their mergers is closer to LF17 ($n^{\mathrm{ISM}}$ = -0.48). After the quenching begins, however, their shape is more similar to that of C00. The epoch marking the coalescence is the one closest to the current state of the Overlap Region. As shown in Figure 6, the average power law exponent for regions from the Overlap is $n^{\mathrm{ISM}}$ = -0.5 which is  very close to the value from LF17, and in close agreement with the results of \cite{roebuck}.

To conclude, we find that our results for the individual regions of Arp 244 are consistent with the trends set by the models of \cite{chevallard} and \cite{trayford}. The same goes for the Overlap Region and the model of \cite{roebuck}. However, the whole system of the Antennae is not dominated by this Overlap Region:  when we estimate the power law exponent of the attenuation curve for Arp 244,  we find a value close the CF00 one. Some high redshift galaxies, such as those observed by ALMA, may share similar properties with an 'extended' Overlap Region.

\section{Summary and conclusions}

We have investigated the reliability of the estimates of physical parameters with  CIGALE by fitting the UV-to-far-IR SEDs of 58 regions of the Antennae Galaxies.
We used the same configuration for the SFH, dust attenuation and re-emission, for all the regions and the galaxy as a whole. We mapped the spatial distribution of the star formation rate and the stellar mass, the stellar mass produced during the last burst, the attenuation in the $V$-band and its power law slope and the abundance of PAHs. Our primary findings are the following:

\begin{itemize}
    \item We compare the estimates of the physical parameters obtained with the 58 regions added to those obtained for Arp 244 as a whole. We find a good agreement showing that the inability to resolve distant objects does not hinder CIGALE estimations.
    \item A general flattening of the attenuation curves with increasing attenuation is found, in agreement with radiative transfer modelling although our measurements correspond to a slightly flatter shape for a given $A_\mathrm{V}^{\mathrm{ISM}}$.
    \item The attenuation curve recipe proposed by \cite{lofaro} for high redshift merger-induced galaxies is the most appropriate to reproduce the SEDs of regions in the Overlap. This compatibility is also consistent with the findings \cite{roebuck} for major mergers up to $z$ \textasciitilde 3.
    \item The trend between the slope of the attenuation curve and the dust surface density is in concordance with the work of \cite{trayford}. Similarly, we found that regions with high dust surface density tend to have a shallower attenuation curve than the CF00 value.
    
\end{itemize}

From our study of Arp 244, we conclude that the use of a SED fitting tool which preserves the energy balance between stellar and dust emission such as CIGALE to measure global physical parameters like  the star formation rate or the stellar mass gives robust results despite the highly inhomogeneous  distributions of UV-optical and dust emission across the system.

\begin{acknowledgements}

We thank the anonymous referee for her/his very useful comments and suggestions. Médéric Boquien gratefully acknowledges support by the ANID BASAL project FB210003 and from the FONDECYT regular grant 1211000.  

\end{acknowledgements}

%
%
\bibliographystyle{aa}
\bibliography{biblio.bib}

\begin{thebibliography}{80}
\expandafter\ifx\csname natexlab\endcsname\relax\def\natexlab#1{#1}\fi

\bibitem[{{Allain} {et~al.}(1996){Allain}, {Leach}, \& {Sedlmayr}}]{allain}
{Allain}, T., {Leach}, S., \& {Sedlmayr}, E. 1996, \aap, 305, 616

\bibitem[{{Baes} {et~al.}(2003){Baes}, {Davies}, {Dejonghe}, {Sabatini},
  {Roberts}, {Evans}, {Linder}, {Smith}, \& {de Blok}}]{baes03}
{Baes}, M., {Davies}, J.~I., {Dejonghe}, H., {et~al.} 2003, \mnras, 343, 1081

\bibitem[{{Baes} {et~al.}(2011){Baes}, {Verstappen}, {De Looze}, {Fritz},
  {Saftly}, {Vidal P{\'e}rez}, {Stalevski}, \& {Valcke}}]{baes11}
{Baes}, M., {Verstappen}, J., {De Looze}, I., {et~al.} 2011, \apjs, 196, 22

\bibitem[{{Battisti} {et~al.}(2017){Battisti}, {Calzetti}, \&
  {Chary}}]{battisti17}
{Battisti}, A.~J., {Calzetti}, D., \& {Chary}, R.~R. 2017, \apj, 851, 90

\bibitem[{{Battisti} {et~al.}(2020){Battisti}, {Cunha}, {Shivaei}, \&
  {Calzetti}}]{battisti20}
{Battisti}, A.~J., {Cunha}, E.~d., {Shivaei}, I., \& {Calzetti}, D. 2020, \apj,
  888, 108

\bibitem[{{Bendo} {et~al.}(2012){Bendo}, {Galliano}, \& {Madden}}]{bendo}
{Bendo}, G.~J., {Galliano}, F., \& {Madden}, S.~C. 2012, \mnras, 423, 197

\bibitem[{{Boquien} {et~al.}(2022){Boquien}, {Buat}, {Burgarella}, {Bardelli},
  {Bethermin}, {Faisst}, {Ginolfi}, {Hathi}, {Jones}, {Koekemoer}, {Lemaux},
  {Narayanan}, {Romano}, {Schaerer}, {Vergani}, {Zamorani}, \&
  {Zucca}}]{boquien22}
{Boquien}, M., {Buat}, V., {Burgarella}, D., {et~al.} 2022, arXiv e-prints,
  arXiv:2202.11723

\bibitem[{{Boquien} {et~al.}(2014){Boquien}, {Buat}, \& {Perret}}]{boquien14}
{Boquien}, M., {Buat}, V., \& {Perret}, V. 2014, \aap, 571, A72

\bibitem[{{Boquien} {et~al.}(2019){Boquien}, {Burgarella}, {Roehlly}, {Buat},
  {Ciesla}, {Corre}, {Inoue}, \& {Salas}}]{boquien19}
{Boquien}, M., {Burgarella}, D., {Roehlly}, Y., {et~al.} 2019, \aap, 622, A103

\bibitem[{{Boselli} {et~al.}(2004){Boselli}, {Lequeux}, \& {Gavazzi}}]{boselli}
{Boselli}, A., {Lequeux}, J., \& {Gavazzi}, G. 2004, \aap, 428, 409

\bibitem[{{Bruzual} \& {Charlot}(2003)}]{bruzual}
{Bruzual}, G. \& {Charlot}, S. 2003, \mnras, 344, 1000

\bibitem[{{Buat} {et~al.}(2018){Buat}, {Boquien}, {Ma{\l}ek}, {Corre}, {Salas},
  {Roehlly}, {Shirley}, \& {Efstathiou}}]{buat18}
{Buat}, V., {Boquien}, M., {Ma{\l}ek}, K., {et~al.} 2018, \aap, 619, A135

\bibitem[{{Buat} {et~al.}(2019){Buat}, {Ciesla}, {Boquien}, {Ma{\l}ek}, \&
  {Burgarella}}]{buat19}
{Buat}, V., {Ciesla}, L., {Boquien}, M., {Ma{\l}ek}, K., \& {Burgarella}, D.
  2019, \aap, 632, A79

\bibitem[{{Buat} {et~al.}(2014){Buat}, {Heinis}, {Boquien}, {Burgarella},
  {Charmandaris}, {Boissier}, {Boselli}, {Le Borgne}, \& {Morrison}}]{buat14}
{Buat}, V., {Heinis}, S., {Boquien}, M., {et~al.} 2014, \aap, 561, A39

\bibitem[{{Buat} {et~al.}(2012){Buat}, {Noll}, {Burgarella}, {Giovannoli},
  {Charmandaris}, {Pannella}, {Hwang}, {Elbaz}, {Dickinson}, {Magdis}, {Reddy},
  \& {Murphy}}]{buat12}
{Buat}, V., {Noll}, S., {Burgarella}, D., {et~al.} 2012, \aap, 545, A141

\bibitem[{{Calzetti} {et~al.}(2000){Calzetti}, {Armus}, {Bohlin}, {Kinney},
  {Koornneef}, \& {Storchi-Bergmann}}]{calzetti}
{Calzetti}, D., {Armus}, L., {Bohlin}, R.~C., {et~al.} 2000, \apj, 533, 682

\bibitem[{{Camps} \& {Baes}(2015)}]{camps15}
{Camps}, P. \& {Baes}, M. 2015, Astronomy and Computing, 9, 20

\bibitem[{{Camps} {et~al.}(2016){Camps}, {Trayford}, {Baes}, {Theuns},
  {Schaller}, \& {Schaye}}]{camps16}
{Camps}, P., {Trayford}, J.~W., {Baes}, M., {et~al.} 2016, \mnras, 462, 1057

\bibitem[{{Cardelli} {et~al.}(1989){Cardelli}, {Clayton}, \&
  {Mathis}}]{cardelli}
{Cardelli}, J.~A., {Clayton}, G.~C., \& {Mathis}, J.~S. 1989, \apj, 345, 245

\bibitem[{{Carnall} {et~al.}(2018){Carnall}, {McLure}, {Dunlop}, \&
  {Dav{\'e}}}]{carnall}
{Carnall}, A.~C., {McLure}, R.~J., {Dunlop}, J.~S., \& {Dav{\'e}}, R. 2018,
  \mnras, 480, 4379

\bibitem[{{Chabrier}(2003)}]{chabrier}
{Chabrier}, G. 2003, \pasp, 115, 763

\bibitem[{{Chambers} {et~al.}(2016){Chambers}, {Magnier}, {Metcalfe},
  {Flewelling}, {Huber}, {Waters}, {Denneau}, {Draper}, {Farrow}, {Finkbeiner},
  {Holmberg}, {Koppenhoefer}, {Price}, {Rest}, {Saglia}, {Schlafly}, {Smartt},
  {Sweeney}, {Wainscoat}, {Burgett}, {Chastel}, {Grav}, {Heasley}, {Hodapp},
  {Jedicke}, {Kaiser}, {Kudritzki}, {Luppino}, {Lupton}, {Monet}, {Morgan},
  {Onaka}, {Shiao}, {Stubbs}, {Tonry}, {White}, {Ba{\~n}ados}, {Bell},
  {Bender}, {Bernard}, {Boegner}, {Boffi}, {Botticella}, {Calamida},
  {Casertano}, {Chen}, {Chen}, {Cole}, {Deacon}, {Frenk}, {Fitzsimmons},
  {Gezari}, {Gibbs}, {Goessl}, {Goggia}, {Gourgue}, {Goldman}, {Grant},
  {Grebel}, {Hambly}, {Hasinger}, {Heavens}, {Heckman}, {Henderson}, {Henning},
  {Holman}, {Hopp}, {Ip}, {Isani}, {Jackson}, {Keyes}, {Koekemoer}, {Kotak},
  {Le}, {Liska}, {Long}, {Lucey}, {Liu}, {Martin}, {Masci}, {McLean}, {Mindel},
  {Misra}, {Morganson}, {Murphy}, {Obaika}, {Narayan}, {Nieto-Santisteban},
  {Norberg}, {Peacock}, {Pier}, {Postman}, {Primak}, {Rae}, {Rai}, {Riess},
  {Riffeser}, {Rix}, {R{\"o}ser}, {Russel}, {Rutz}, {Schilbach}, {Schultz},
  {Scolnic}, {Strolger}, {Szalay}, {Seitz}, {Small}, {Smith}, {Soderblom},
  {Taylor}, {Thomson}, {Taylor}, {Thakar}, {Thiel}, {Thilker}, {Unger},
  {Urata}, {Valenti}, {Wagner}, {Walder}, {Walter}, {Watters}, {Werner},
  {Wood-Vasey}, \& {Wyse}}]{chambers}
{Chambers}, K.~C., {Magnier}, E.~A., {Metcalfe}, N., {et~al.} 2016, arXiv
  e-prints, arXiv:1612.05560

\bibitem[{{Charlot} \& {Fall}(2000)}]{charlot}
{Charlot}, S. \& {Fall}, S.~M. 2000, \apj, 539, 718

\bibitem[{{Chevallard} {et~al.}(2013){Chevallard}, {Charlot}, {Wandelt}, \&
  {Wild}}]{chevallard}
{Chevallard}, J., {Charlot}, S., {Wandelt}, B., \& {Wild}, V. 2013, \mnras,
  432, 2061

\bibitem[{{Ciesla} {et~al.}(2015){Ciesla}, {Charmandaris}, {Georgakakis},
  {Bernhard}, {Mitchell}, {Buat}, {Elbaz}, {LeFloc'h}, {Lacey}, {Magdis}, \&
  {Xilouris}}]{ciesla}
{Ciesla}, L., {Charmandaris}, V., {Georgakakis}, A., {et~al.} 2015, \aap, 576,
  A10

\bibitem[{{Corre} {et~al.}(2018){Corre}, {Buat}, {Basa}, {Boissier}, {Japelj},
  {Palmerio}, {Salvaterra}, {Vergani}, \& {Zafar}}]{corre}
{Corre}, D., {Buat}, V., {Basa}, S., {et~al.} 2018, \aap, 617, A141

\bibitem[{{Crain} {et~al.}(2015){Crain}, {Schaye}, {Bower}, {Furlong},
  {Schaller}, {Theuns}, {Dalla Vecchia}, {Frenk}, {McCarthy}, {Helly},
  {Jenkins}, {Rosas-Guevara}, {White}, \& {Trayford}}]{crain}
{Crain}, R.~A., {Schaye}, J., {Bower}, R.~G., {et~al.} 2015, \mnras, 450, 1937

\bibitem[{{da Cunha} {et~al.}(2008){da Cunha}, {Charlot}, \& {Elbaz}}]{dacunha}
{da Cunha}, E., {Charlot}, S., \& {Elbaz}, D. 2008, \mnras, 388, 1595

\bibitem[{{Draine} {et~al.}(2014){Draine}, {Aniano}, {Krause}, {Groves},
  {Sandstrom}, {Braun}, {Leroy}, {Klaas}, {Linz}, {Rix}, {Schinnerer},
  {Schmiedeke}, \& {Walter}}]{draine14}
{Draine}, B.~T., {Aniano}, G., {Krause}, O., {et~al.} 2014, \apj, 780, 172

\bibitem[{{Draine} \& {Li}(2007)}]{draine07}
{Draine}, B.~T. \& {Li}, A. 2007, \apj, 657, 810

\bibitem[{{Dunlop} {et~al.}(2017){Dunlop}, {McLure}, {Biggs}, {Geach},
  {Micha{\l}owski}, {Ivison}, {Rujopakarn}, {van Kampen}, {Kirkpatrick},
  {Pope}, {Scott}, {Swinbank}, {Targett}, {Aretxaga}, {Austermann}, {Best},
  {Bruce}, {Chapin}, {Charlot}, {Cirasuolo}, {Coppin}, {Ellis}, {Finkelstein},
  {Hayward}, {Hughes}, {Ibar}, {Jagannathan}, {Khochfar}, {Koprowski},
  {Narayanan}, {Nyland}, {Papovich}, {Peacock}, {Rieke}, {Robertson},
  {Vernstrom}, {Werf}, {Wilson}, \& {Yun}}]{dunlop}
{Dunlop}, J.~S., {McLure}, R.~J., {Biggs}, A.~D., {et~al.} 2017, \mnras, 466,
  861

\bibitem[{{Elbaz} {et~al.}(2018){Elbaz}, {Leiton}, {Nagar}, {Okumura},
  {Franco}, {Schreiber}, {Pannella}, {Wang}, {Dickinson}, {D{\'\i}az-Santos},
  {Ciesla}, {Daddi}, {Bournaud}, {Magdis}, {Zhou}, \& {Rujopakarn}}]{elbaz}
{Elbaz}, D., {Leiton}, R., {Nagar}, N., {et~al.} 2018, \aap, 616, A110

\bibitem[{{Fazio} {et~al.}(2004){Fazio}, {Hora}, {Allen}, {Ashby}, {Barmby},
  {Deutsch}, {Huang}, {Kleiner}, {Marengo}, {Megeath}, {Melnick}, {Pahre},
  {Patten}, {Polizotti}, {Smith}, {Taylor}, {Wang}, {Willner}, {Hoffmann},
  {Pipher}, {Forrest}, {McMurty}, {McCreight}, {McKelvey}, {McMurray}, {Koch},
  {Moseley}, {Arendt}, {Mentzell}, {Marx}, {Losch}, {Mayman}, {Eichhorn},
  {Krebs}, {Jhabvala}, {Gezari}, {Fixsen}, {Flores}, {Shakoorzadeh}, {Jungo},
  {Hakun}, {Workman}, {Karpati}, {Kichak}, {Whitley}, {Mann}, {Tollestrup},
  {Eisenhardt}, {Stern}, {Gorjian}, {Bhattacharya}, {Carey}, {Nelson},
  {Glaccum}, {Lacy}, {Lowrance}, {Laine}, {Reach}, {Stauffer}, {Surace},
  {Wilson}, {Wright}, {Hoffman}, {Domingo}, \& {Cohen}}]{fazio}
{Fazio}, G.~G., {Hora}, J.~L., {Allen}, L.~E., {et~al.} 2004, \apjs, 154, 10

\bibitem[{{Feldmann} {et~al.}(2017){Feldmann}, {Quataert}, {Hopkins},
  {Faucher-Gigu{\`e}re}, \& {Kere{\v{s}}}}]{feldmann}
{Feldmann}, R., {Quataert}, E., {Hopkins}, P.~F., {Faucher-Gigu{\`e}re}, C.-A.,
  \& {Kere{\v{s}}}, D. 2017, \mnras, 470, 1050

\bibitem[{{Fontanot} {et~al.}(2009){Fontanot}, {Somerville}, {Silva}, {Monaco},
  \& {Skibba}}]{fontanot}
{Fontanot}, F., {Somerville}, R.~S., {Silva}, L., {Monaco}, P., \& {Skibba}, R.
  2009, \mnras, 392, 553

\bibitem[{{Gil de Paz} {et~al.}(2007){Gil de Paz}, {Boissier}, {Madore},
  {Seibert}, {Joe}, {Boselli}, {Wyder}, {Thilker}, {Bianchi}, {Rey}, {Rich},
  {Barlow}, {Conrow}, {Forster}, {Friedman}, {Martin}, {Morrissey}, {Neff},
  {Schiminovich}, {Small}, {Donas}, {Heckman}, {Lee}, {Milliard}, {Szalay}, \&
  {Yi}}]{gil}
{Gil de Paz}, A., {Boissier}, S., {Madore}, B.~F., {et~al.} 2007, \apjs, 173,
  185

\bibitem[{{G{\'o}mez-Guijarro} {et~al.}(2018){G{\'o}mez-Guijarro}, {Toft},
  {Karim}, {Magnelli}, {Magdis}, {Jim{\'e}nez-Andrade}, {Capak}, {Fraternali},
  {Fujimoto}, {Riechers}, {Schinnerer}, {Smol{\v{c}}i{\'c}}, {Aravena},
  {Bertoldi}, {Cortzen}, {Hasinger}, {Hu}, {Jones}, {Koekemoer}, {Lee},
  {McCracken}, {Micha{\l}owski}, {Navarrete}, {Povi{\'c}}, {Puglisi},
  {Romano-D{\'\i}az}, {Sheth}, {Silverman}, {Staguhn}, {Steinhardt},
  {Stockmann}, {Tanaka}, {Valentino}, {van Kampen}, \& {Zirm}}]{gomez}
{G{\'o}mez-Guijarro}, C., {Toft}, S., {Karim}, A., {et~al.} 2018, \apj, 856,
  121

\bibitem[{{Gonzalez-Perez} {et~al.}(2013){Gonzalez-Perez}, {Lacey}, {Baugh},
  {Frenk}, \& {Wilkins}}]{gonzalez}
{Gonzalez-Perez}, V., {Lacey}, C.~G., {Baugh}, C.~M., {Frenk}, C.~S., \&
  {Wilkins}, S.~M. 2013, \mnras, 429, 1609

\bibitem[{{Hodge} {et~al.}(2016){Hodge}, {Swinbank}, {Simpson}, {Smail},
  {Walter}, {Alexander}, {Bertoldi}, {Biggs}, {Brandt}, {Chapman}, {Chen},
  {Coppin}, {Cox}, {Dannerbauer}, {Edge}, {Greve}, {Ivison}, {Karim},
  {Knudsen}, {Menten}, {Rix}, {Schinnerer}, {Wardlow}, {Weiss}, \& {van der
  Werf}}]{hodge}
{Hodge}, J.~A., {Swinbank}, A.~M., {Simpson}, J.~M., {et~al.} 2016, \apj, 833,
  103

\bibitem[{{Jonsson}(2006)}]{jonsson06}
{Jonsson}, P. 2006, \mnras, 372, 2

\bibitem[{{Jonsson} {et~al.}(2010){Jonsson}, {Groves}, \& {Cox}}]{jonsson10}
{Jonsson}, P., {Groves}, B.~A., \& {Cox}, T.~J. 2010, \mnras, 403, 17

\bibitem[{{Karl} {et~al.}(2013){Karl}, {Lunttila}, {Naab}, {Johansson},
  {Klaas}, \& {Juvela}}]{karl13}
{Karl}, S.~J., {Lunttila}, T., {Naab}, T., {et~al.} 2013, \mnras, 434, 696

\bibitem[{{Karl} {et~al.}(2010){Karl}, {Naab}, {Johansson}, {Kotarba}, {Boily},
  {Renaud}, \& {Theis}}]{karl10}
{Karl}, S.~J., {Naab}, T., {Johansson}, P.~H., {et~al.} 2010, \apjl, 715, L88

\bibitem[{{Kennicutt} \& {Evans}(2012)}]{kennicutt}
{Kennicutt}, R.~C. \& {Evans}, N.~J. 2012, \araa, 50, 531

\bibitem[{{Klaas} {et~al.}(2010){Klaas}, {Nielbock}, {Haas}, {Krause}, \&
  {Schreiber}}]{klaas}
{Klaas}, U., {Nielbock}, M., {Haas}, M., {Krause}, O., \& {Schreiber}, J. 2010,
  \aap, 518, L44

\bibitem[{{Kriek} \& {Conroy}(2013)}]{kriek}
{Kriek}, M. \& {Conroy}, C. 2013, \apjl, 775, L16

\bibitem[{{Lah{\'e}n} {et~al.}(2018){Lah{\'e}n}, {Johansson}, {Rantala},
  {Naab}, \& {Frigo}}]{lahen}
{Lah{\'e}n}, N., {Johansson}, P.~H., {Rantala}, A., {Naab}, T., \& {Frigo}, M.
  2018, \mnras, 475, 3934

\bibitem[{{Lo Faro} {et~al.}(2017){Lo Faro}, {Buat}, {Roehlly},
  {Alvarez-Marquez}, {Burgarella}, {Silva}, \& {Efstathiou}}]{lofaro}
{Lo Faro}, B., {Buat}, V., {Roehlly}, Y., {et~al.} 2017, \mnras, 472, 1372

\bibitem[{{Magnier} {et~al.}(2020){Magnier}, {Schlafly}, {Finkbeiner}, {Tonry},
  {Goldman}, {R{\"o}ser}, {Schilbach}, {Casertano}, {Chambers}, {Flewelling},
  {Huber}, {Price}, {Sweeney}, {Waters}, {Denneau}, {Draper}, {Hodapp},
  {Jedicke}, {Kaiser}, {Kudritzki}, {Metcalfe}, {Stubbs}, \&
  {Wainscoat}}]{magnier}
{Magnier}, E.~A., {Schlafly}, E.~F., {Finkbeiner}, D.~P., {et~al.} 2020, \apjs,
  251, 6

\bibitem[{{Ma{\l}ek} {et~al.}(2018){Ma{\l}ek}, {Buat}, {Roehlly}, {Burgarella},
  {Hurley}, {Shirley}, {Duncan}, {Efstathiou}, {Papadopoulos}, {Vaccari},
  {Farrah}, {Marchetti}, \& {Oliver}}]{malek}
{Ma{\l}ek}, K., {Buat}, V., {Roehlly}, Y., {et~al.} 2018, \aap, 620, A50

\bibitem[{{McMahon} {et~al.}(2013){McMahon}, {Banerji}, {Gonzalez}, {Koposov},
  {Bejar}, {Lodieu}, {Rebolo}, \& {VHS Collaboration}}]{mcmahon}
{McMahon}, R.~G., {Banerji}, M., {Gonzalez}, E., {et~al.} 2013, The Messenger,
  154, 35

\bibitem[{{Mirabel} {et~al.}(1998){Mirabel}, {Vigroux}, {Charmandaris},
  {Sauvage}, {Gallais}, {Tran}, {Cesarsky}, {Madden}, \& {Duc}}]{mirabel}
{Mirabel}, I.~F., {Vigroux}, L., {Charmandaris}, V., {et~al.} 1998, \aap, 333,
  L1

\bibitem[{{Noll} {et~al.}(2009){Noll}, {Burgarella}, {Giovannoli}, {Buat},
  {Marcillac}, \& {Mu{\~n}oz-Mateos}}]{noll}
{Noll}, S., {Burgarella}, D., {Giovannoli}, E., {et~al.} 2009, \aap, 507, 1793

\bibitem[{{Pantoni} {et~al.}(2021){Pantoni}, {Lapi}, {Massardi}, {Donevski},
  {Bressan}, {Silva}, {Pozzi}, {Vignali}, {Talia}, {Cimatti}, {Ronconi}, \&
  {Danese}}]{pantoni}
{Pantoni}, L., {Lapi}, A., {Massardi}, M., {et~al.} 2021, \mnras, 504, 928

\bibitem[{{Pierini} {et~al.}(2004){Pierini}, {Gordon}, {Witt}, \&
  {Madsen}}]{pierini}
{Pierini}, D., {Gordon}, K.~D., {Witt}, A.~N., \& {Madsen}, G.~J. 2004, \apj,
  617, 1022

\bibitem[{{Pilbratt} {et~al.}(2010){Pilbratt}, {Riedinger}, {Passvogel},
  {Crone}, {Doyle}, {Gageur}, {Heras}, {Jewell}, {Metcalfe}, {Ott}, \&
  {Schmidt}}]{pilbratt}
{Pilbratt}, G.~L., {Riedinger}, J.~R., {Passvogel}, T., {et~al.} 2010, \aap,
  518, L1

\bibitem[{{Privon} {et~al.}(2013){Privon}, {Barnes}, {Evans}, {Hibbard}, {Yun},
  {Mazzarella}, {Armus}, \& {Surace}}]{privon}
{Privon}, G.~C., {Barnes}, J.~E., {Evans}, A.~S., {et~al.} 2013, \apj, 771, 120

\bibitem[{{Renaud} {et~al.}(2015){Renaud}, {Bournaud}, \& {Duc}}]{renaud}
{Renaud}, F., {Bournaud}, F., \& {Duc}, P.-A. 2015, \mnras, 446, 2038

\bibitem[{{Rieke} {et~al.}(2004){Rieke}, {Young}, {Engelbracht}, {Kelly},
  {Low}, {Haller}, {Beeman}, {Gordon}, {Stansberry}, {Misselt}, {Cadien},
  {Morrison}, {Rivlis}, {Latter}, {Noriega-Crespo}, {Padgett}, {Stapelfeldt},
  {Hines}, {Egami}, {Muzerolle}, {Alonso-Herrero}, {Blaylock}, {Dole}, {Hinz},
  {Le Floc'h}, {Papovich}, {P{\'e}rez-Gonz{\'a}lez}, {Smith}, {Su}, {Bennett},
  {Frayer}, {Henderson}, {Lu}, {Masci}, {Pesenson}, {Rebull}, {Rho}, {Keene},
  {Stolovy}, {Wachter}, {Wheaton}, {Werner}, \& {Richards}}]{rieke}
{Rieke}, G.~H., {Young}, E.~T., {Engelbracht}, C.~W., {et~al.} 2004, \apjs,
  154, 25

\bibitem[{{Riess} {et~al.}(2011){Riess}, {Macri}, {Casertano}, {Lampeitl},
  {Ferguson}, {Filippenko}, {Jha}, {Li}, \& {Chornock}}]{riess}
{Riess}, A.~G., {Macri}, L., {Casertano}, S., {et~al.} 2011, \apj, 730, 119

\bibitem[{{Rodriguez-Gomez} {et~al.}(2019){Rodriguez-Gomez}, {Snyder}, {Lotz},
  {Nelson}, {Pillepich}, {Springel}, {Genel}, {Weinberger}, {Tacchella},
  {Pakmor}, {Torrey}, {Marinacci}, {Vogelsberger}, {Hernquist}, \&
  {Thilker}}]{rodriguez}
{Rodriguez-Gomez}, V., {Snyder}, G.~F., {Lotz}, J.~M., {et~al.} 2019, \mnras,
  483, 4140

\bibitem[{{Roebuck} {et~al.}(2019){Roebuck}, {Sajina}, {Hayward}, {Martis},
  {Marchesini}, {Krefting}, \& {Pope}}]{roebuck}
{Roebuck}, E., {Sajina}, A., {Hayward}, C.~C., {et~al.} 2019, \apj, 881, 18

\bibitem[{{Roussel}(2013)}]{roussel}
{Roussel}, H. 2013, \pasp, 125, 1126

\bibitem[{{Rujopakarn} {et~al.}(2019){Rujopakarn}, {Daddi}, {Rieke}, {Puglisi},
  {Schramm}, {P{\'e}rez-Gonz{\'a}lez}, {Magdis}, {Alberts}, {Bournaud},
  {Elbaz}, {Franco}, {Kawinwanichakij}, {Kohno}, {Narayanan}, {Silverman},
  {Wang}, \& {Williams}}]{rujopakarn19}
{Rujopakarn}, W., {Daddi}, E., {Rieke}, G.~H., {et~al.} 2019, \apj, 882, 107

\bibitem[{{Rujopakarn} {et~al.}(2016){Rujopakarn}, {Dunlop}, {Rieke}, {Ivison},
  {Cibinel}, {Nyland}, {Jagannathan}, {Silverman}, {Alexander}, {Biggs},
  {Bhatnagar}, {Ballantyne}, {Dickinson}, {Elbaz}, {Geach}, {Hayward},
  {Kirkpatrick}, {McLure}, {Micha{\l}owski}, {Miller}, {Narayanan}, {Owen},
  {Pannella}, {Papovich}, {Pope}, {Rau}, {Robertson}, {Scott}, {Swinbank}, {van
  der Werf}, {van Kampen}, {Weiner}, \& {Windhorst}}]{rujopakarn16}
{Rujopakarn}, W., {Dunlop}, J.~S., {Rieke}, G.~H., {et~al.} 2016, \apj, 833, 12

\bibitem[{{Saftly} {et~al.}(2015){Saftly}, {Baes}, {De Geyter}, {Camps},
  {Renaud}, {Guedes}, \& {De Looze}}]{saftly}
{Saftly}, W., {Baes}, M., {De Geyter}, G., {et~al.} 2015, \aap, 576, A31

\bibitem[{{Salim} {et~al.}(2018){Salim}, {Boquien}, \& {Lee}}]{salim18}
{Salim}, S., {Boquien}, M., \& {Lee}, J.~C. 2018, \apj, 859, 11

\bibitem[{{Salim} \& {Narayanan}(2020)}]{salim20}
{Salim}, S. \& {Narayanan}, D. 2020, \araa, 58, 529

\bibitem[{{Salmon} {et~al.}(2016){Salmon}, {Papovich}, {Long}, {Willner},
  {Finkelstein}, {Ferguson}, {Dickinson}, {Duncan}, {Faber}, {Hathi},
  {Koekemoer}, {Kurczynski}, {Newman}, {Pacifici}, {P{\'e}rez-Gonz{\'a}lez}, \&
  {Pforr}}]{salmon}
{Salmon}, B., {Papovich}, C., {Long}, J., {et~al.} 2016, \apj, 827, 20

\bibitem[{{Sanders} {et~al.}(2003){Sanders}, {Mazzarella}, {Kim}, {Surace}, \&
  {Soifer}}]{sanders}
{Sanders}, D.~B., {Mazzarella}, J.~M., {Kim}, D.~C., {Surace}, J.~A., \&
  {Soifer}, B.~T. 2003, \aj, 126, 1607

\bibitem[{{Schaye} {et~al.}(2015){Schaye}, {Crain}, {Bower}, {Furlong},
  {Schaller}, {Theuns}, {Dalla Vecchia}, {Frenk}, {McCarthy}, {Helly},
  {Jenkins}, {Rosas-Guevara}, {White}, {Baes}, {Booth}, {Camps}, {Navarro},
  {Qu}, {Rahmati}, {Sawala}, {Thomas}, \& {Trayford}}]{schaye}
{Schaye}, J., {Crain}, R.~A., {Bower}, R.~G., {et~al.} 2015, \mnras, 446, 521

\bibitem[{{Schlafly} \& {Finkbeiner}(2011)}]{schlafly}
{Schlafly}, E.~F. \& {Finkbeiner}, D.~P. 2011, \apj, 737, 103

\bibitem[{{Silva} {et~al.}(1998){Silva}, {Granato}, {Bressan}, \&
  {Danese}}]{silva}
{Silva}, L., {Granato}, G.~L., {Bressan}, A., \& {Danese}, L. 1998, \apj, 509,
  103

\bibitem[{{Springel}(2005)}]{springel}
{Springel}, V. 2005, \mnras, 364, 1105

\bibitem[{{Tasca} {et~al.}(2015){Tasca}, {Le F{\`e}vre}, {Hathi}, {Schaerer},
  {Ilbert}, {Zamorani}, {Lemaux}, {Cassata}, {Garilli}, {Le Brun}, {Maccagni},
  {Pentericci}, {Thomas}, {Vanzella}, {Zucca}, {Amorin}, {Bardelli},
  {Cassar{\`a}}, {Castellano}, {Cimatti}, {Cucciati}, {Durkalec}, {Fontana},
  {Giavalisco}, {Grazian}, {Paltani}, {Ribeiro}, {Scodeggio}, {Sommariva},
  {Talia}, {Tresse}, {Vergani}, {Capak}, {Charlot}, {Contini}, {de la Torre},
  {Dunlop}, {Fotopoulou}, {Koekemoer}, {L{\'o}pez-Sanjuan}, {Mellier}, {Pforr},
  {Salvato}, {Scoville}, {Taniguchi}, \& {Wang}}]{tasca}
{Tasca}, L.~A.~M., {Le F{\`e}vre}, O., {Hathi}, N.~P., {et~al.} 2015, \aap,
  581, A54

\bibitem[{{Trayford} {et~al.}(2020){Trayford}, {Lagos}, {Robotham}, \&
  {Obreschkow}}]{trayford}
{Trayford}, J.~W., {Lagos}, C. d.~P., {Robotham}, A. S.~G., \& {Obreschkow}, D.
  2020, \mnras, 491, 3937

\bibitem[{{Tuffs} {et~al.}(2004){Tuffs}, {Popescu}, {V{\"o}lk}, {Kylafis}, \&
  {Dopita}}]{tuffs}
{Tuffs}, R.~J., {Popescu}, C.~C., {V{\"o}lk}, H.~J., {Kylafis}, N.~D., \&
  {Dopita}, M.~A. 2004, \aap, 419, 821

\bibitem[{{Villa-V{\'e}lez} {et~al.}(2021){Villa-V{\'e}lez}, {Buat},
  {Theul{\'e}}, {Boquien}, \& {Burgarella}}]{villa}
{Villa-V{\'e}lez}, J.~A., {Buat}, V., {Theul{\'e}}, P., {Boquien}, M., \&
  {Burgarella}, D. 2021, \aap, 654, A153

\bibitem[{{Whitmore} \& {Schweizer}(1995)}]{Whitmore}
{Whitmore}, B.~C. \& {Schweizer}, F. 1995, \aj, 109, 960

\bibitem[{{Zhang} {et~al.}(2010){Zhang}, {Gao}, \& {Kong}}]{zhang}
{Zhang}, H.-X., {Gao}, Y., \& {Kong}, X. 2010, \mnras, 401, 1839

\end{thebibliography}

\begin{appendix}

\section{Background noise map}

In figure A.1 we present the locus of the 66 regions used to compute the background noise in all the images used in this work and listed in Table 1.

\begin{figure}[hbt!]
\centering
\includegraphics[width=0.9\paperwidth]{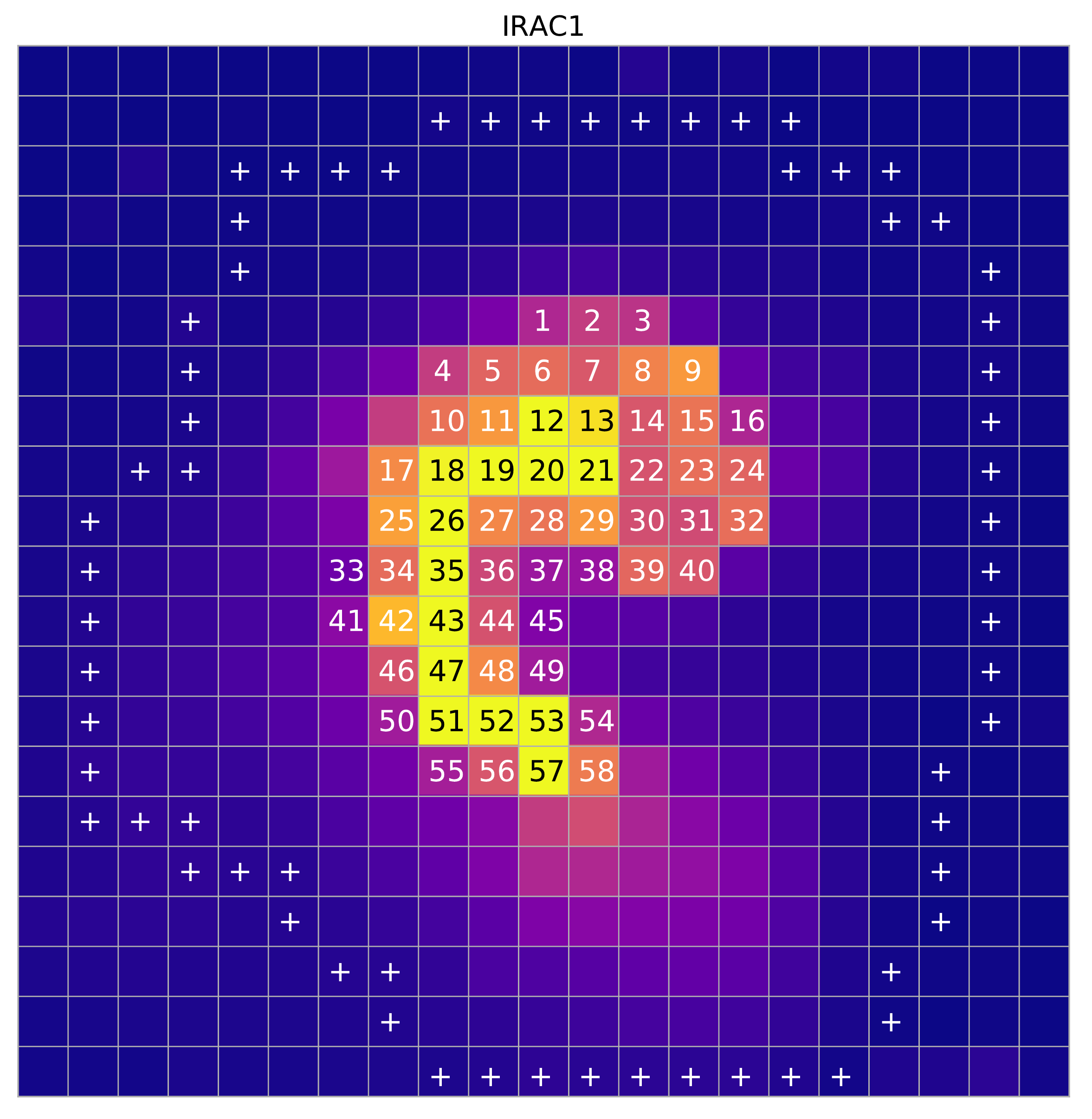}
\caption{ The 66 regions (marked with white crosses) are represented on the IRAC1 band image.}
\end{figure}

\end{appendix}

\end{document}